\documentclass{aa}  
\usepackage{graphicx}
\usepackage{natbib}
\setcitestyle{aysep={}} 
\usepackage{hyperref}
\hypersetup{
    colorlinks = true,
    linkcolor = {blue}, 
    citecolor = {blue},
    urlcolor = {blue}
}

\usepackage{longtable}
\usepackage{lscape}
\usepackage{placeins}
\usepackage{txfonts}
\begin{document}

   \title{Deep high-resolution L band spectroscopy in the $\beta$~Pictoris planetary system\thanks{Based on observations from the European Southern Observatory, Chile (Programme 0110.C-4301(A)).}
}


   \author{Markus Janson\inst{1} \and
          Jonas Wehrung-Montpezat\inst{1,2} \and
          Ansgar Wehrhahn\inst{1} \and
          Alexis Brandeker\inst{1} \and
          Gayathri Viswanath\inst{1} \and
          Paul Molli{\`e}re\inst{3} \and
          Thomas Stolker\inst{4}
          }

   \institute{Department of Astronomy, Stockholm University, AlbaNova University Center, 10691 Stockholm, Sweden\\
              \email{markus.janson@astro.su.se}
        \and
        {\'E}cole Polytechnique, Institut Polytechnique de Paris, 91120 Palaiseau, France
        \and
        Max-Planck-Institut f\"{u}r Astronomie, K\"{o}nigstuhl 17, D-69117 Heidelberg, Germany
        \and
        Leiden Observatory, Leiden University, Niels Bohrweg 2, 2333 CA Leiden, The Netherlands
        }

   \date{Received ---; accepted ---}

   \abstract{
   The $\beta$~Pictoris system, with its two directly imaged planets $\beta$~Pic b and $\beta$~Pic c and its well characterised debris disk, is a prime target for detailed characterisation of young planetary systems. Here, we present high-resolution and high-contrast LM band spectroscopy with CRIRES+ of the system, primarily for the purpose of atmospheric characterisation of $\beta$~Pic b. We developed methods for determining slit geometry and wavelength calibration based on telluric absorption and emission lines, as well as methods for point spread function (PSF) modelling and subtraction, and artificial planet injection, in order to extract and characterise planet spectra at a high signal-to-noise ratio ($S/N$) and spectral fidelity. Through cross-correlation with model spectra, we detected H$_2$O absorption for planet b in each of the 13 individual observations spanning four different spectral settings. This provides a clear confirmation of previously detected water absorption, and allowed us to derive an exquisite precision on the rotational velocity of $\beta$~Pic b, $v_{\rm rot} = 20.36 \pm 0.31$ km/s, which is consistent within error bars with previous determinations. We also observed a tentative H$_2$O cross-correlation peak at the expected position and velocity of planet c; the feature is however not at a statistically significant level. Despite a higher sensitivity to SiO than earlier studies, we do not confirm a tentative SiO feature previously reported for planet b. When combining data from different epochs and different observing modes for the strong H$_2$O feature of planet b, we find that the $S/N$ grows considerably faster when sets of different spectral settings are combined, compared to when multiple data sets of the same spectral setting are combined. This implies that maximising spectral coverage is often more important than maximising integration depth when investigating exoplanetary atmospheres using cross-correlation techniques.
   }


\titlerunning{$\beta$~Pic L band planetary spectroscopy}
\authorrunning{M. Janson et al.}

   \maketitle
%

\section{Introduction}
\label{s:intro}

With an age of 24$\pm$3 Myr \citep{bell2015} and a distance of only 19.7 pc \citep{gaia2016}, the A6V \cite{gray2006} star $\beta$~Pictoris is one of the best-suited stars in the sky for detailed studies of a young circumstellar environment. Additionally, its circumstellar disk \citep{aumann1985} is extraordinarily bright and large, extending out to at least 2000 au \citep{janson2021}, which led to it becoming the first disk ever outside of the Solar System to be directly imaged \citep{smith1984}. The disk shows an interesting set of sub-structure and features \citep{kalas1995}, including a warp due to an inclined secondary disk \citep[e.g.][]{heap2000,golimowski2006}; a very large density of comets \citep{lecavelier2022,janson2023} with a range of eccentricities going all the way up to star-grazing orbits \citep{ferlet1987,beust1990}; and the presence of gasses with short expected lifetimes in the strongly irradiated disk, implying recent gas release in large planetesimal disruptions \citep{brandeker2016,cataldi2018}. All these features indicate the presence of planets that dynamically excite various components of the disk. The secondary disk, in particular, has been used for rather precise predictions of a planet orbiting within the disk.

In 2010, the planet $\beta$~Pic~b was indeed confirmed in the system through direct imaging \citep{lagrange2010}, when common proper motion and orbital motion were confirmed for a candidate that had previously been discovered in a first epoch through reprocessing of archival images \citep{lagrange2009}. The imaged planet was well consistent with the planet that had been predicted on the basis of the orientation of the secondary disk, which originates from the slight misalignment between the planes of the main disk component and the planetary orbit \citep{chauvin2012}. Later on, a second planet, $\beta$~Pic~c, was identified first through radial velocity (RV) measurements \citep{lagrange2019} and soon after confirmed through direct imaging with the interferometer GRAVITY \citep{nowak2020}, further demonstrating that $\beta$~Pic is a rich and dynamic planetary system. The favourable observational properties of the system have led to exquisite orbital determinations of the planets through (primarily) GRAVITY interferometry, to the point that it has been possible to constrain the mass of planet c directly through its impact on the astrometry of planet b \citep{lacour2021}. The planet masses and semi-major axes are estimated in the ranges of 7--12~$M_{\rm jup}$ and $\sim$10 au for b, and 7--10~$M_{\rm jup}$ and $\sim$2.7 au for c \citep{brandt2021,lacour2021}.

Much like in the case of astrometric characterisation, the favourable observational properties of the $\beta$~Pic system make its planets particularly well suited for spectroscopic characterisation -- particularly planet b, due to its larger semi-major axis, leading to a higher angular separation from the star in most epochs. This led to $\beta$~Pic~b being the first exoplanet to be successfully studied with high-resolution spectroscopy\footnote{Several of the HR~8799 planets had previously been studied at a lower spectral resolution \citep[e.g.][]{janson2010,konopacky2013}}\citep{snellen2014}. The high-resolution ($R$ $\sim$ 50\,000--100\,000) spectroscopy was acquired in the K band with the CRyogenic high-resolution InfraRed Echelle Spectrograph \citep[CRIRES;][]{kaufl2004,dorn2023} at the Very Large Telescope (VLT) managed by the European Southern Observatories (ESO). The principle behind the technique is based on cross-correlation function (CCF) techniques for effectively summing up the information from whole forests of lines at once, in order to account for the fact that each individual spectral line typically has a very low signal-to-noise ratio ($S/N$) in the measured spectra. The resulting cross-correlation peak retains many of the features of the individual lines it is based on, and can therefore be used to derive kinematic properties such as the rotational velocity and RV of the planet. Cross-correlation has also been successfully shown to work for medium-resolution ($R$ $\sim$5000) spectroscopy for $\beta$~Pic~b \citep{hoeijmakers2018,kiefer2024}. These techniques are well complemented by spectroscopic studies at a lower resolution that can provide constraints on the absolute flux level as a function of wavelength, which is typically challenging to derive from an analysis based on cross-correlation. Recent results to this end include James Webb Space Telescope (JWST) Mid-InfraRed Imager (MIRI) spectroscopy of $\beta$~Pic b at 5--7 $\mu$m \citep{worthen2024}, and GRAVITY K band spectroscopy of both planets b and c \citep{gravity2020,nowak2020}.

Following an upgrade of CRIRES (named CRIRES+) which greatly widens its simultaneous spectral range and enhances its sensitivity, renewed attention has been placed on characterising $\beta$~Pic~b at a high spectral resolution. Recent studies have investigated the planet in the K band \citep{landman2024} and in the M band \citep{parker2024}. Both studies have reported detections of H$_2$O and CO. The M band study also noted a tentative detection of SiO. In between the K and M bands lies the L band, which as of yet has been less studied spectroscopically. Yet, it is an important wavelength range for the study of young planets, for several reasons. As many direct imaging and spectroscopic studies have noted \citep[e.g.][]{kasper2007,janson2010,quanz2010,mawet2013,janson2015}, the wavelength range around $\sim$4~$\mu$m is advantageous in a high-contrast context, due to the favourable contrast between stars and planets for a wide range of planetary temperatures \citep[e.g.][]{burrows2006,linder2019} in combination with a relatively low thermal background noise in the L band compared to longer wavelengths, as well as the high adaptive optics (AO) stability that can be achieved in the L band. Furthermore, the L band contains strong opacity regions of molecules that are expected to be important in planetary atmospheres, including H$_2$O, CH$_4$, and NH$_3$, and more exotic species such as SiO and SO$_2$. 

In this paper, we present CRIRES+ L band spectroscopy of the $\beta$~Pic planetary system. While the main target is $\beta$~Pic b, the CRIRES+ slit alignment automatically also includes $\beta$~Pic c, and we therefore attempted to extract signal from c as well. The L band range is particularly well suited for studying a contrast-limited planet such as c that would be very challenging to reach at shorter wavelengths, although as we subsequently see, the observational conditions for c were not optimal during these observations. The paper is structured as follows: In Sect. \ref{s:obs}, we describe the observations acquired for the purpose of this study. The reduction of the collected data is outlined in Sect. \ref{s:data}, and the subsequent analysis and corresponding results follow in Sect. \ref{s:results}. We discuss the outcomes of the study and consequences for future studies in Sect. \ref{s:discussion}, and finally briefly summarise the conclusions in Sect. \ref{s:summary}.

\section{Observations}
\label{s:obs}

Observations were carried out with CRIRES+ in Service Mode starting on 29 Nov 2022. Originally scheduled for ESO observing period P110, only about half of the observations were carried out in that period, the rest being carried over into P111 with the last observation carried out during 1 Nov 2023. We used four different spectral settings in order to cover all of the L band with minimum gaps and overlaps, and parts of the M band. The observations were divided into 15 observing blocks (OBs), with 3 OBs each dedicated to settings L3262, L3340 and L3426, and 6 OBs dedicated to setting M4318. One incomplete OB executed on 26 Feb 2023 failed validation and was later re-executed; we have discarded the failed version of the OB. The full set of validated observations is listed in Table~\ref{t:log}, where on each of Oct 12 and 13, two OBs have been merged and are treated as a single observation, since they were acquired consecutively and in the same mode. 

\begin{table*}[htb]
\caption{Observing log for the $\beta$~Pic CRIRES+ observations.}
\label{t:log}
\centering
\begin{tabular}{llllllllll}
\hline
\hline
Date & Mode & DIT & NDIT & $N_{\rm fr}$ & $\rho_{\rm b}$ & RV$_{\rm b}$ & $\rho_{\rm c}$ & RV$_{\rm c}$ & BV$^{a}$ \\
 &  &  & (s) & & (mas) & (km/s) & (mas) & (km/s) & (km/s) \\
\hline
29 Nov 2022     &       L3262   &       30      &       1       &       120     &       536     &       11.0$\pm$0.3    &       11      &       -0.4$\pm$3.9    &       2.44    \\
23 Dec 2022     &       L3262   &       30      &       1       &       120     &       539     &       11.1$\pm$0.3    &       23      &       -6.0$\pm$3.8    &       -0.95   \\
23 Dec 2022     &       M4318   &       10      &       6       &       52      &       539     &       11.1$\pm$0.3    &       23      &       -6.0$\pm$3.8    &       -0.95   \\
25 Dec 2022     &       L3340   &       30      &       1       &       120     &       539     &       11.1$\pm$0.3    &       29      &       -6.4$\pm$3.8    &       -1.23   \\
31 Dec 2022     &       L3426   &       60      &       1       &       60      &       539     &       11.1$\pm$0.3    &       32      &       -7.6$\pm$3.7    &       -2.08   \\
22 Jan 2023     &       M4318   &       10      &       6       &       56      &       541     &       11.1$\pm$0.3    &       54      &       -11.7$\pm$3.3   &       -4.91   \\
15 Feb 2023     &       L3340   &       60      &       1       &       60      &       544     &       11.2$\pm$0.3    &       72      &       -15.2$\pm$2.9   &       -7.13   \\
25 Feb 2023     &       M4318   &       10      &       6       &       56      &       544     &       11.2$\pm$0.3    &       81      &       -16.4$\pm$2.6   &       -7.70   \\      
27 Sep 2023     &       L3340   &       30      &       1       &       128     &       557     &       11.4$\pm$0.3    &       149     &       -20.2$\pm$0.5   &       7.94    \\
9 Oct 2023      &       L3262   &       60      &       1       &       60      &       557     &       11.4$\pm$0.3    &       149     &       -19.8$\pm$0.5   &       7.52    \\
12 Oct 2023$^{b}$       &       L3426   &       60      &       1       &       124     &       557     &       11.4$\pm$0.3    &       149     &       -19.7$\pm$0.5   &       7.36    \\
13 Oct 2023$^{b}$       &       M4318   &       10      &       6       &       104     &       557     &       11.4$\pm$0.3    &       149     &       -19.6$\pm$0.5   &       7.30    \\
1 Nov 2023      &       M4318   &       10      &       6       &       56      &       558     &       11.4$\pm$0.3    &       147     &       -18.9$\pm$0.5   &       5.81    \\
\hline
\end{tabular}
\begin{list}{}{}
\item[$^{\mathrm{a}}$] $N_{\rm fr}$: Number of frames in the run. $\rho$: Star-planet separation. RV$_{\rm b}$ and RV$_{\rm c}$: Expected planet RV relative to the star for planets b and c. BV: Barycentric velocity of Earth relative to $\beta$~Pic.
\item[$^{\mathrm{b}}$] Two OBs in the same mode were executed contiguously on these nights, the OB pairs are merged and count as single observations here.
\end{list}
\end{table*}

In the same table, we also list the expected separation and RV for planets $\beta$~Pic b and c at each respective epoch, based on orbital parameters in \citet{lacour2021}. Likewise, the barycentric RV of the Earth relative to $\beta$~Pic is listed for each epoch. Due to the long baseline between first and last epoch, there is significant evolution in these quantities from epoch to epoch, which we take into account when combining the data in Sect. \ref{s:cc}. The $\beta$~Pic system RV relative to the barycenter of the Solar System is 20.0 km/s \citep{gaia2023}. For all observations, the position angle of the slit was 31.55 deg, which comfortably encompasses the full orbital planes of planets b and c in the nearly edge-on $\beta$~Pic system. The slit width was consistently 200 mas, leading to a spectral resolution of $\sim$100\,000. Planet b was our main target, but since c is also included at all times, we consider it a secondary target. Unfortunately, the observing epochs were not ideal for c, as we will return to in Sect. \ref{s:planetc}. However, the epochs were (by design) well suited for detection and characterisation of planet b, which resided close to its maximum combined separation and RV relative to the star across the observational baseline.

Observations were acquired using an ABBA nodding pattern in order to enable pairwise subtraction of the high sky level in L band observations. We implemented the metrology procedure offered by ESO at the beginning of each observation in order to ensure reproducibility of the spectral positioning on the detectors between different nights. For all L band settings (L3262, L3340, and L3426), our standard setting was a direct integration time (DIT) of 60 seconds per exposure, with two exposures per nodding position over 15 nodding cycles. For some runs, this was manually adjusted to DIT of 30 s and four exposures per nodding position by ESO staff, in cases where they estimated an elevated risk of saturation. The total on-source integration time per L band observation was therefore 1 hour exactly. For the OBs with the M band setting (M4318), we set DIT to 10 s and the number of integrations per exposure (NDIT) to six with two exposures per nodding position over 13--14 nodding cycles (half with 13 and half with 14) for on-source integration times of 52--56 minutes per OB. 

Every spectral setting used for the observations includes six spectral orders, which are each divided over three adjacent detectors in CRIRES+. The data in each epoch can therefore be seen as consisting of 18 spectral regions, which we here refer to simply as `chunks'. For all reduction and analysis steps up until the cross-correlation discussed in Sect. \ref{s:cc}, we operate on each chunk individually. This means that we are able to operate on continuous data sets of $\sim$30 nm width each, in which the point spread function (PSF) is not expected to significantly change from one edge wavelength to the other. For the M band setting M4318, two orders (six chunks) were excluded for each frame: One of the orders overlaps with the fundamental band of telluric CO$_2$ where the transmission is effectively zero across the whole order, and the second excluded order was the longest-wavelength order in which the thermal background was high enough that the combined star and background flux saturated the detector. The chunks were reduced and analysed as described in the following sections.  

\section{Data reduction and extraction}
\label{s:data}

\subsection{Reduction and calibration}
\label{s:calib}

CRIRES+ has a dedicated data reduction pipeline based on the PyReduce software \citep{piskunov2021}, which under standard circumstances can automatically accomplish the entire data reduction chain from raw data through to fully extracted and calibrated spectra for point source targets. However, in our case, there are two main reasons for why the automatic procedure cannot be directly applied to the data: Firstly, our primary objective is not to extract just the spectrum of the star $\beta$~Pic A, which dominates the signal in a point-source extraction, but rather to subtract the signal of A and subsequently extract the signals of the fainter b and c companions -- this procedure is discussed in Sect.~\ref{s:psf} and Sect.~\ref{s:extraction}. Secondly, and of more immediate relevance to the early phases of data reduction, ESO provides no lamp spectra or similar types of calibration files for L or M band observations, which the pipeline needs for its slit tilt, slit curvature, and wavelength calibration steps. We therefore devised a reduction scheme that uses the routines of the standard pipeline whenever possible, with custom procedures mixed in for all steps where this was necessary. 

In this way, we produced frames that were corrected for bias, dark, and flatfield effects based on the raw data and standard calibration files. These `cleaned' frames were used for both custom calibration and science extraction purposes, with different processing for each branch. As outlined above, the purpose of the custom calibration in this context was the determination of slit tilt/curvature and wavelength calibration. Both of these calibrations are usually based on lamp spectra, where the full slit is illuminated by emission lines from a known compound. The morphology of the lines on the detector can then be used to determine the orientation and curvature of the slit image as function of wavelength, and the positions of the lines with known emitted wavelengths can be measured to provide the wavelength calibration. In the absence of such lamp spectra, we devised two separate methods for achieving these calibrations based on the science data. 

In one method, we used the telluric absorption lines in the $\beta$~Pic stellar spectra. For each absorption line, the A and B nodding positions provided two data points suitable for determining the slit tilt at that particular wavelength. In principle, the slit image also has a mild curvature in addition to the tilt, which cannot be measured with only two points -- however, this curvature is essentially negligible \citep[see][]{piskunov2021}, so we disregard it in this analysis. A second degree polynomial was then fit to the degree of tilt as function of wavelength, in order to interpolate (or in the case of near-edge pixels, extrapolate) the expected tilt in regions that lacked telluric absorption lines. For wavelength calibration, we then used PySME \citep{wehrhahn2023} for identifying the telluric line species and thereby construct a wavelength solution along each spectral trace. 

In the second method, we used telluric emission lines instead of absorption lines. For this purpose, we created `artificial' sky frames by dividing every spectral order along the slit direction (near-vertical on the detector) into a lower and upper half. For each A and B nodding frame, we then saved only the half that did not contain the star, put the remaining halves together to create sky frames where both the upper and lower parts were devoid of starlight, and averaged the results into deep frames containing only (to first order) light from the sky. Due to the high image depth of the observations, this provided sky frames with a reasonably rich set of lines in most orders. Slit tilts and wavelength calibrations were then determined in an analogous way as in the absorption-based method. The two methods complemented each other well, in that some regions that were relatively poor in absorption lines still contained a decent number of emission lines, and vice versa. In orders where both methods produced acceptable results, those results were mutually consistent, with the absorption method generally producing higher-quality spectral lines and thus being the preferred choice in such cases.

With this procedure, we were able to produce slit geometry and wavelength calibration, in addition to the standard calibration produced in the default manner. The standard pipeline procedure could then be used to produce a fully reduced and extracted spectrum of the star $\beta$~Pic, if interpreted as a single point-source. This is useful for normalisation purposes, as we will see in Sect.~\ref{s:psf}. However, the main scientific objectives of the project require us to perform PSF subtraction on the star $\beta$~Pic A and subsequently extract the signals of the much fainter planets $\beta$~Pic b and c. That procedure will be discussed in the following section. The customised procedure for determining slit geometry and wavelength calibration described above, for the spectral settings we used (L3262, L3340, L3426, and M4318), is available on github\footnote{\url{https://github.com/AWehrhahn/CRIRES_LM}}.

\subsection{PSF determination and subtraction}
\label{s:psf}

The PSF subtraction scheme performed here is a variant of the STARK routine previously implemented on VLT/UVES \citep{ringqvist2023} and JWST/NIRCAM \citep{patel2024} data, which is available on github\footnote{\url{https://github.com/Jayshil/stark}}. For the PSF subtraction purposes, we operate on the $A-B$ and $B-A$ pairwise background-subtracted images produced in the procedure described in the previous section. The input pixel values are fluxes in units of detector counts, and each value is also associated with a corresponding error estimation based on the calculated photon count (from non-background subtracted frames) and read noise in each pixel. The main features of the procedure are described below.

The CRIRES+ spectra are mapped on the detector(s) in such a way that the dispersion (wavelength) direction is primarily horizontal and the spatial direction along the slit is primarily vertical. However, as discussed in Sect.~\ref{s:calib}, the slit image is in fact not completely vertical, but has a slight tilt, with a tilt angle that changes slowly across the wavelength axis. In other words, pixels along a detector column (within an order) correspond to similar, but not equal, wavelengths. Thus, as a first step, we consider each pixel within a spectral chunk with pixel indices $x_i$ for the row number and $y_i$ for the column number on the detector. Based on the wavelength and slit tilt calibrations described previously, we calculate the pixel's wavelength $\lambda_i$ and its spatial separation along the slit from the stellar trace, $\rho_i$. 

\begin{figure}[htb]
\centering
\includegraphics[width=9.2cm]{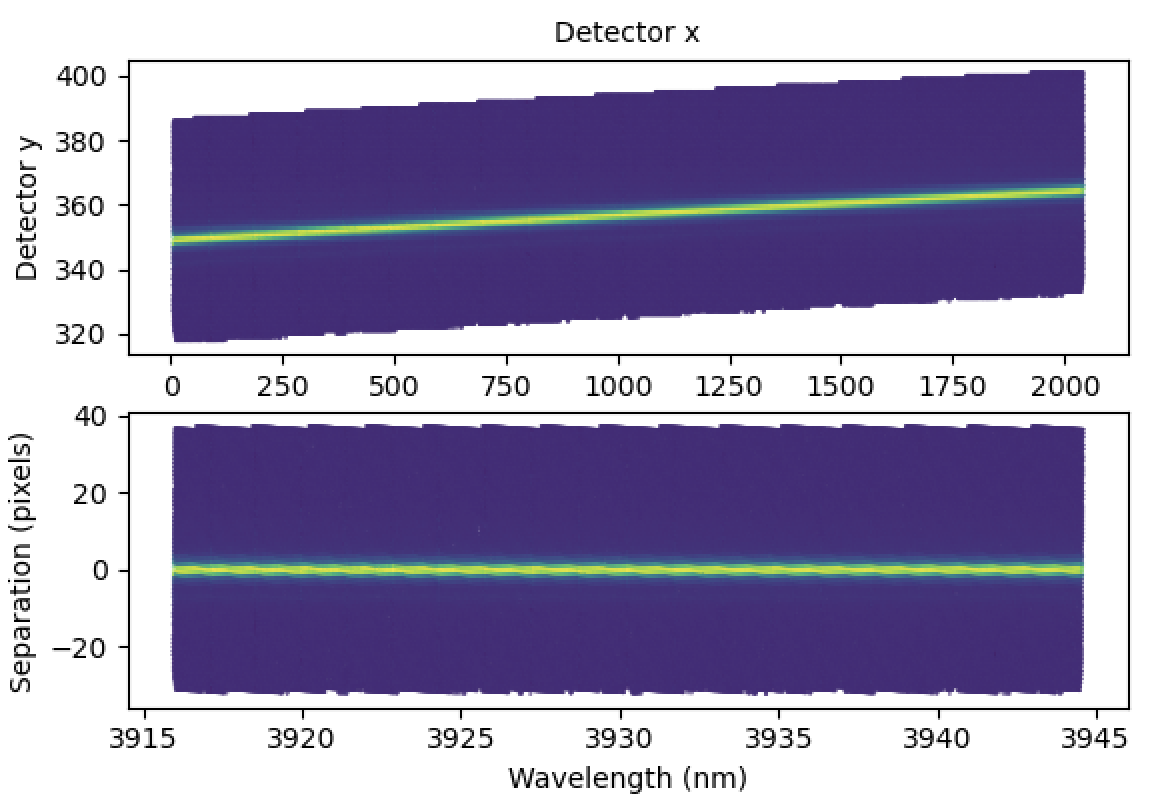}
\caption{Flux-normalised spectral traces of the star $\beta$~Pic, for one single spectral chunk. The panels are not plotted as pixeled images (which would not be possible for the bottom panel, since it is not uniformly sampled), but as one coloured dot per sampled data point. Yellow is a normalised flux of 1, blue is 0. Top panel: Flux sampled by the $x_i$ and $y_i$ coordinates on the detector. Bottom panel: The same flux (no interpolation) sampled by wavelength $\lambda_i$ and separation $\rho_i$.}
\label{f:xyrl}
\end{figure}

At this point, all relevant pixels are assigned wavelengths and distances from a central coordinate, which as a collective make up a non-uniformly sampled grid. An intuitively natural next step might then be to interpolate a new spectral image, which is sampled uniformly in wavelength-separation space. However, operating with interpolated data at this stage of the process can negatively impact the contrast performance, since imperfections in the interpolation can lead to large errors for any faint planets residing in the bright stellar PSF wings. Instead, we represent the PSF by using a $\chi^2$ minimisation fit to the data with a model in the form of a 2D spline in wavelength-separation space. A spline does not need to operate on a uniform grid, and can therefore be applied directly to the non-interpolated $\lambda_i$ and $\rho_i$ coordinates. This procedure also makes it easier to account for hot/bad pixels, which can be identified through sigma clipping and simply ignored in the fitting. To simplify the fitting as much as possible, we renormalise the data by wavelength prior to the fit using the previously extracted stellar 1d-spectrum. This straightens out both tellurics and stellar features, creating a function that always peaks at a value of 1 at the centre of the PSF for all wavelengths. Importantly, when re-scaling the fluxes, we also re-scale their corresponding error estimates by the same factor. The error estimates are included in the $\chi^2$ fitting, and mean (for example) that data points inside of deep telluric lines acquire a very low weight in the fitting, since the relative error is very large inside such features. 

For the spline fitting, we use a uniform grid of knots with $\sim$600 knot points along the spatial ($\rho_i$) direction, and only 3 knot points along the spectral ($\lambda_i$) direction. The specific number of spatial samples is arbitrary, but is set to be high in order to provide an accurate PSF fit along the spatial axis, whilst still maintaining the total number of knot points much smaller than the total number of available data points ($\sim$120,000) per chunk. Meanwhile, the reason for the modest spectral sampling is to allow the PSF evolve, but only very smoothly, along the spectral axis. This is sufficient to characterise chromatic PSF effects (which should be small and smooth, given the short wavelength span of the spectral chunks), but simultaneously, it minimises damaging impact on the planetary spectral signal. The continuum flux of any planet around the star inevitably gets included in the PSF fit to the star, and is therefore subtracted out in the PSF subtraction; however, planetary spectral lines are much narrower than the spacing between the spectral spline knots, and thus impossible for the spline to fit. Such lines are therefore efficiently conserved when the fit is subtracted. This principle is confirmed by our injection testing in Sect.~\ref{s:injection}. 

\begin{figure}[htb]
\centering
\includegraphics[width=9.2cm]{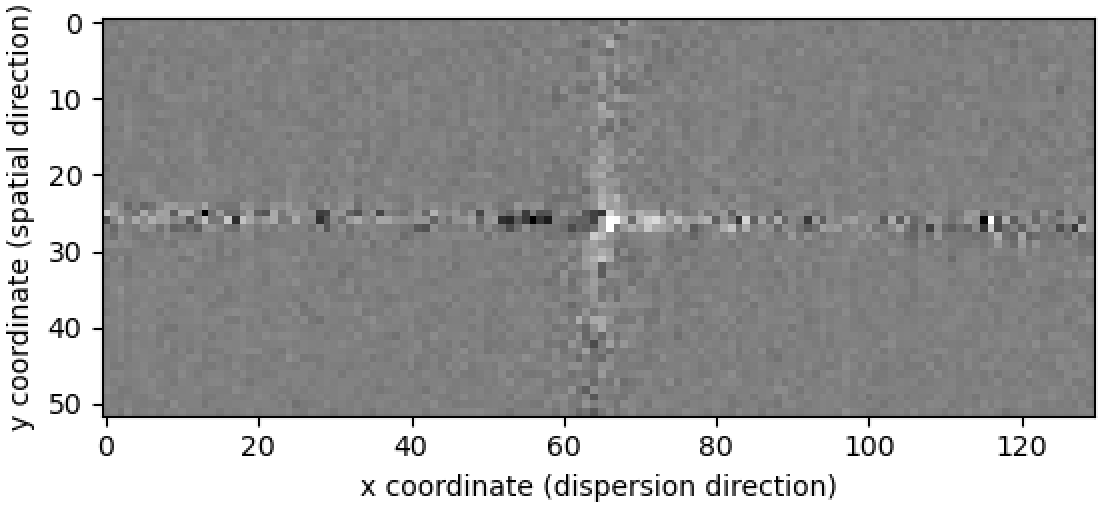}
\caption{Example of residual data after PSF subtraction. The figure is in detector xy-space, since it can then be plotted as a non-interpolated pixeled image. Spatially (vertically), the image is centred on the stellar trace, which after PSF subtraction only remains as a narrow horizontal region of increased noise. Spectrally (horizontally), the image is centred on a strong telluric absorption feature. In this particular region, the PSF subtraction does not perform as well as along the rest of the spectrum, resulting in a vertical net positive feature extending out from the centre. At even larger separations vertically from the centre, the noise in the telluric region is generally enhanced. This feature would remain even if the PSF subtraction had been perfect, since it represents the much lower transmission inside of the telluric line than in the continuum. Planet b is $\sim$9.5 pixels below the stellar trace, but cannot be seen without cross-correlation techniques.}
\label{f:res}
\end{figure}

In this way, it is possible to produce a 2D spectral PSF model, evaluate it at the exact $\lambda_i$ and $\rho_i$ points sampled on the detector, and subtract the model from the data, without any need to interpolate the data itself. The result is a residual frame that contains only noise at the location of the star, but retains signal (in the form of spectral lines) from all planets around the star, although such signals are too weak in each individual pixel to be distinguishable by eye. As can be seen in Fig.~\ref{f:res}, the procedure works well except in particular spectral regions, which correspond exactly to where there are strong telluric lines in the spectra. Effectively, the PSF as mapped on the detector appears to have a slightly different shape inside and outside of the tellurics. The reason for this is either a slight non-linearity in the detector readout, or a consequence of the fact that the PSF is slightly narrower than the slit, leading to a slightly non-uniform slit illumination \citep[see][]{landman2024}. There are ways to account for these effects that we expect to implement in later versions of the procedure, but for the purpose of this study, the telluric regions are in any case not usable, due to the low transmission and consequential high relative noise in those regions. Instead, we simply let the telluric regions get masked out during the sigma clipping in Sect.~\ref{s:extraction}. 

\subsection{Extraction of planetary spectra}
\label{s:extraction}

From the residual frame after stellar PSF subtraction, we can extract spectra from the circumstellar region in different ways. A straightforward example is to extract the spectra of planets b and c based on a priori knowledge about their locations relative to the star from previous astrometry \citep[][]{lacour2021}. For doing so, we define synthetic `sub-slits' centred on the known exoplanet location, with a slit length of 5 pixels along the spatial direction (i.e. 2.5 pixels in each direction from the planet centre). The extraction is done epoch by epoch, so the slit location varies with the planetary motion over time. This is not particularly important for planet b, since its motion is not very large ($\sim$0.5 pixels) over the observational baseline, but considerably more important for planet c with its much more rapid motion. 

The sub-slit now represents an aperture that contains the vast majority of the planetary flux. From this, we extract a 1D spectrum with a wavelength sampling corresponding to one pixel on the detector. Since the $\lambda_i$ and $\rho_i$ coordinate grid is non-uniform relative to the pixel grid, there is not an equal number of data points in every spectral bin, and the data points are not equally distributed along the planetary PSF, so a straight aperture sum of the flux would not work satisfactorily. Instead, we implement a PSF-normalised weighted aperture sum to account for these effects. For this purpose, we use a shifted version of the already determined stellar PSF from the PSF subtraction procedure, to now represent the planetary PSF. Every data point inside the sub-slit is normalised by $1/f(\rho_p)$ where $f(\rho_p)$ is the flux of the PSF (relative to the peak flux) at separation $\rho_p$ from the planetary PSF centre. We normalise the estimated errors by the same factor. An error-weighted mean is then taken inside the spatial aperture for each spectral window to represent the extracted 1D spectral data points. 

\begin{figure*}[htb]
\centering
\includegraphics[width=18.5cm]{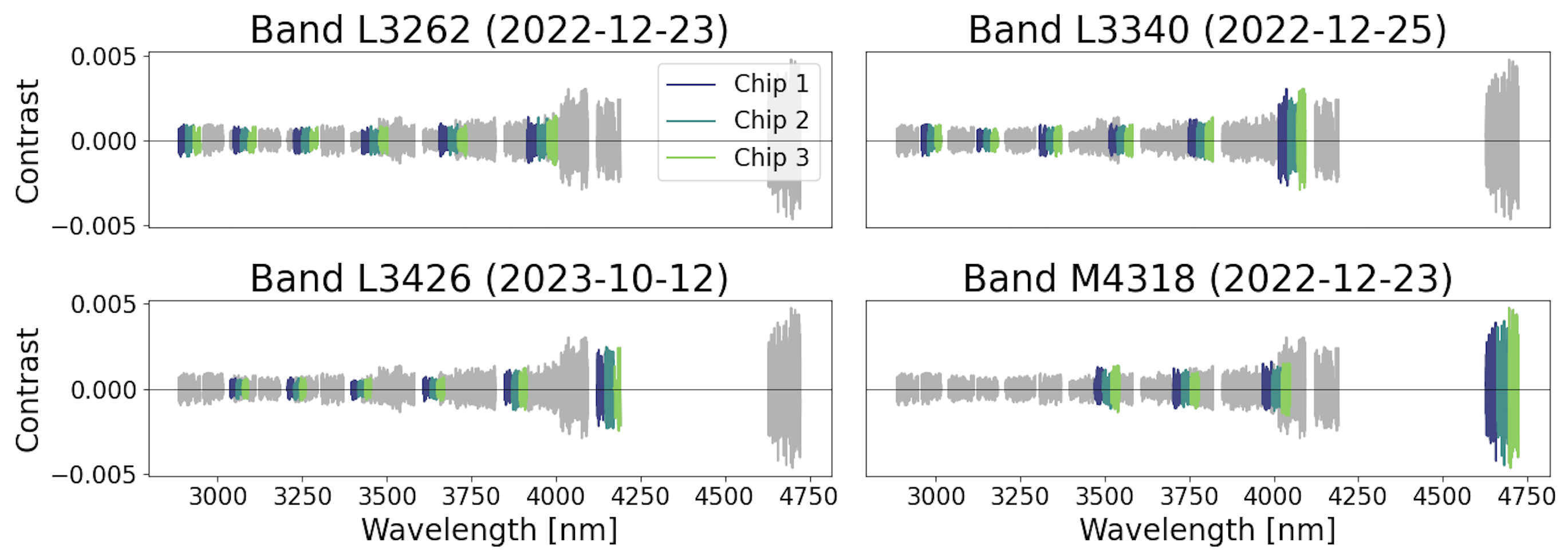}
\caption{Spectral coverage of the observations in this study. Four spectral settings were used: L3262 (upper left), L3340 (upper right), L3426 (lower left) and M4318 (lower right). The coloured regions show the coverage of that particular setting, while the grey regions show the coverage of the other settings. The scatter corresponds to the noise level in each spectral region. Higher scatter can be caused by shorter integration time, worse ambient conditions, or lower instrumental transmission, but here it primarily reflects wavelength ranges of higher thermal background, and/or wavelength ranges of lower atmospheric transmission.}
\label{f:modes}
\end{figure*}

As references for the planetary spectra and for evaluating any possible systematic noise sources affecting them, we also extract `mirror' spectra from the same separations as the planets, but on the exact opposite side of the star. This yields data sets with similar noise properties as the planetary spectra, but without the planetary signals. Furthermore, we perform a series of extractions in the same way as above, but consecutively centred on integer pixel separations from -15 to 15 pixels from the star. This creates a spectral map which can be used for purposes such as blind searches, or checking systematic noise levels. We subject all extracted spectra to an iterative sigma clipping procedure in which $\sigma$ is determined from the scatter in the data, after which $>$4$\sigma$ deviating data points are removed, followed by a new estimation of $\sigma$ until convergence is reached. This efficiently removes heavily telluric-affected regions, as well as remaining bad pixels. Due to the normalisations earlier in the procedure, the spectra are conveniently dimensionless, and correspond to the contrast between the (real or hypothetical) planetary flux and the stellar flux at each sampled wavelength. The spectral coverage and the scatter in the contrast at different wavelengths is illustrated in Fig.~\ref{f:modes}.

\subsection{Injection of artificial planet signals}
\label{s:injection}

For the purpose of the sensitivity evaluation analysis that will be discussed in Sect.~\ref{s:planetb}, we make tests where we inject artificial planet spectra into the data before PSF subtraction and extract them after PSF subtraction. For example, if we wish to evaluate detectability of CO$_2$ in planet b, we first generate a theoretical model spectrum with the expected physical properties of planet b (see Sect.~\ref{s:models} for a description of how the model spectra are generated), and a CO$_2$ atmosphere. We transform the flux spectrum into a contrast spectrum by dividing it with a stellar model spectrum for $\beta$~Pic A. In the PSF determination and subtraction scheme (Sect.~\ref{s:psf}), after the first estimation of the stellar PSF, we then produce an artificial planet by scaling the stellar PSF with the planetary contrast spectrum, and adding the resulting signature into the data at the empty mirror location for $\beta$~Pic b. We then re-estimate the stellar PSF based on the new data, subtract it, and then extract the signal at the location where the artificial planet was introduced. This results in spectra with noise that is well representative of the noise affecting the real planet, and with a controlled artificial CO$_2$ abundance for testing whether that abundance should be detectable or not. The process can then be repeated for other molecules for planet b, and in principle also planet c, although such an analysis for c is more complex, as discussed in Sect.~\ref{s:planetc}.

The above procedure is useful, but quite time consuming, since every new molecule and every new parameter for each molecule requires a new full PSF subtraction procedure to generate artificial contrast spectra. We therefore also implement a 'pseudo-injection' scheme, where instead of adding an artificial planet to the 2D spectra prior to PSF subtraction, we calculate a model contrast spectrum for a given molecule and a given set of parameters for that molecule (e.g.\ temperature, volume mixing ratio, see Sect.~\ref{s:models}). This model contrast is then added to the extracted spectrum on the opposite side of the star from the planet, in order to include a representative noise level. A normal cross-correlation procedure as discussed above is then performed in order to evaluate the detection strength for the (pseudo)-injected signal. The advantage of this method over actual injection is that it is much faster, and can therefore be applied to a wider range of molecules and molecular parameters. The disadvantage is that it fails to include any self-subtraction effects that take place on real signals during the PSF subtraction step. However, as we will see, these effects are quite minor for the existing data sets, so after running a real injection on water and comparing to a pseudo-injection (see Sect.~\ref{s:planetb}) with the same parameters to verify sufficient accuracy, we used the pseudo-injection scheme for most of the injection based analysis of this study.

\section{Analysis and results}
\label{s:results}

The extracted spectra are analysed using cross-correlation techniques, which are frequently employed in high-resolution exoplanet spectroscopy \citep[e.g.][]{snellen2014,schwarz2016,bryan2018,bryan2020,xuan2020}. Here we describe the implementation of the technique, and present the results it has yielded.

\subsection{Spectral models}
\label{s:models}

For spectral modelling, we use \textit{petitRADTRANS} \citep{molliere2019,molliere2020} models based on the relatively well constrained physical properties of planets b and c, with an atmosphere consisting of hydrogen and helium (which contribute atmospheric pressure, but no strong lines in the wavelength region we consider here) plus any individual molecule we intend to search for in the extracted spectra. 

Following \citep{guillot2010} and the previous recent CRIRES+ spectroscopic studies of $\beta$~Pic, we use a temperature-pressure profile assuming negligible irradiation contribution, as follows:

\begin{equation}
T^4 = \left( \frac{1}{2} + \frac{3 P \kappa_{\rm IR}}{4 g} \right) T_{\rm int}^{4}
\end{equation}

where $T$ is the temperature profile as function of pressure $P$, $\kappa_{\rm IR}$ is the infrared absorption coefficient which we set to $\kappa_{\rm IR} = 0.005$~cm$^2$g$^{-1}$, $g$ the surface gravity of the planet, and $T_{\rm int}$ is the internal temperature.

For planet b, we use an internal temperature of 1700 K as in \citet{parker2024}, and $\log g = 4.18$. For planet c, we set the internal temperature to 1250 K based on \citet{nowak2020}, and $\log g = 4.24$. We produce spectra both with and without rotational broadening. The rotational velocity is 20.36 km/s (see Sect.~\ref{s:planetb}) for planet b. For planet c, the corresponding velocity is unknown so we set it to a similar value as for b, 20 km/s. For molecule identification in the extracted spectra, we typically use broadened spectra unless stated otherwise. The results based on non-broadened spectra are very similar. We also fit and subtract the continuum from those spectra, in order to distill only the line information. For the synthetic planet injection described in Sect.~\ref{s:injection}, we use broadened spectra. We also include the instrumental broadening, although it is practically negligible relative to the rotation -- at $R \sim 100\,000$, the instrumental full width at half maximum (FWHM) is approximately 3 km/s.

\subsection{Cross-correlation method}
\label{s:cc}

Through cross-correlation of the extracted spectra with the models from Sect.~\ref{s:models}, we can search for specific molecules by noting which modelled species produce a significant cross-correlation peak; and from the width and velocity shift of any such peaks, we can also determine the rotational velocity and RV of the planet. However, the spectral extraction procedure produces many different spectral chunks from different parts of different spectral orders at different epochs, where each individual chunk is relatively short ($\sim$30 nm) and therefore does not necessarily contain sufficient abundances of lines by itself to yield a significant cross-correlation peak. The true wealth of this data set stems from the fact that it contains 204 different spectral chunks (3 chunks per order, 4--6 orders per spectral setting, 4 spectral settings with 2--5 epochs per setting). For the cross-correlations, we combine the various triplets of chunks corresponding to the same order into continuous spectral regions, but that still leaves 68 spectral regions that need to be combined in some way in order to maximise the detection significance and quality of the results. The fundamental procedure to accomplish this is to cross-correlate each spectral chunk with a corresponding model, and average the resulting cross-correlation functions to enhance the signal-to-noise ratio $S/N$. 

However, taking a straight average over all spectral regions would not be a good path towards this aim, because for a given molecular species, some regions may contain many intrinsically strong lines, while other regions may contain essentially nothing. Mixing empty regions in with the rich regions would then only add noise and no signal, decreasing the total $S/N$. Instead we can, at least in principle, achieve a near-optimal $S/N$ by appropriate weighting during averaging of the different regions. We implement this through a weighted average scheme in which the weights are determined by the expected information content in each region. As proxy for the information content, we use the average spectral gradient in the data. That is to say for each spectral region, we take a derivative of the modelled spectrum in that region, and calculate the mean of its absolute values. This accounts for the variations in the spectrum imposed by the spectral lines: If there are many strong spectral lines within the region, it acquires a correspondingly high weight, while if the there are no lines, the weight is zero and the region is effectively excluded.

We have additionally tried other methods of weighting and combination of weightings, including weighting for the expected signal-to-noise ratio of each spectral region, and using only binary weighting that either includes a region if it contains spectral lines or excludes the region if it is devoid of lines. However, we find that the different weighting methods make very little difference for the outcomes, hence it appears that the specific weighting method is of little importance -- the dominant factor for providing a strong $S/N$ in the cross-correlation output for our current data set is simply the inclusion of data that contains lines, and exclusion of data that contains no lines.

\subsection{Planet b}
\label{s:planetb}

\begin{figure*}[htb]
\centering
\includegraphics[width=15cm]{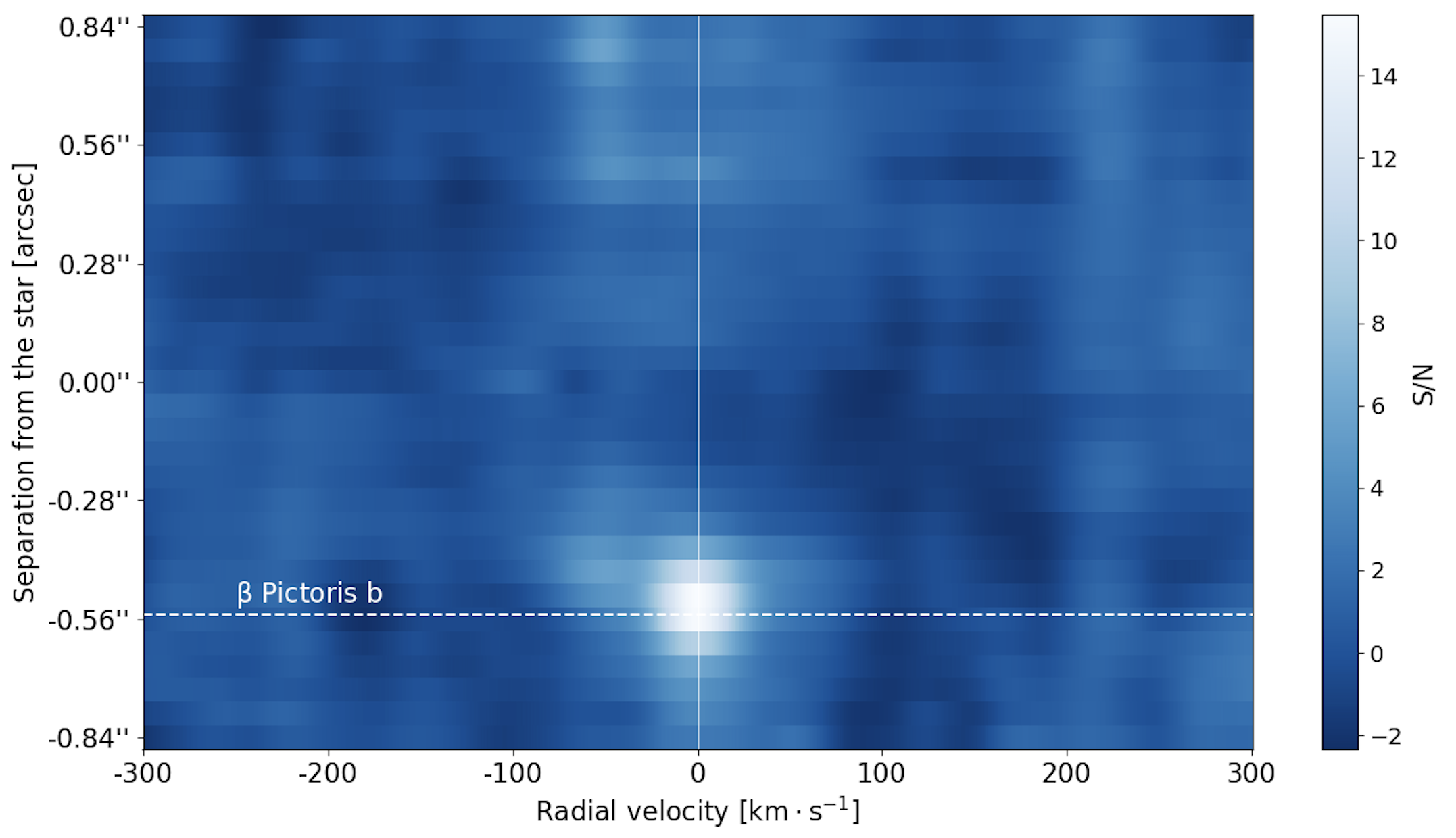}
\caption{Cross-correlation $S/N$ map for H$_2$O, showing a clear detection of $\beta$~Pic b at its expected separation and velocity. The velocity is in the frame of reference of the planet itself. The separation changes between 536 mas and 558 mas during the observational baseline between 29 Nov 2022  and 1 Nov 2023.}
\label{f:ccfmap}
\end{figure*}

The by far most prominent spectral feature of planet b, as observed in the L band, is H$_2$O absorption. In Fig.~\ref{f:ccfmap}, we show a cross-correlation map of the inner $\beta$~Pic system, in the velocity frame of $\beta$~Pic b. While the planet moves in separation by 22 mas (approximately 0.5 pixels) between the first and last epoch of observations, causing some mild vertical smearing in the image, planet b stands out clearly at its expected position with a high level of statistical significance, $S/N = 15.0$. The $S/N$ is evaluated as the cross-correlation peak value divided by the standard deviation across the cc function at the planetary separation, excluding a range of $\pm$70 km/s centred on the planetary velocity. This is the same definition for the $S/N$ metric as used in \citet{parker2024}, and slightly different from the definition in \citet{landman2024}, which evaluates the noise on the basis of the standard deviation of the cross-correlation function at the symmetrically opposite location of the star, relative to the planet. We assess that the stellar PSF in our data set is sufficiently asymmetric that the noise properties at equal separations in opposite directions from the star are not necessarily fully mutually representative, hence we opt for the former approach. The CCF is quite clean from systematic noise in the continuum around the peak, relative to much of the CCF analysis in the literature. This is most likely a consequence of the fact that the LM band data consists of a wide range of different spectral regions, which help in mitigating such systematic noise, as we discuss in Sect. \ref{s:discussion}.

\begin{figure*}[htb]
\centering
\includegraphics[width=15cm]{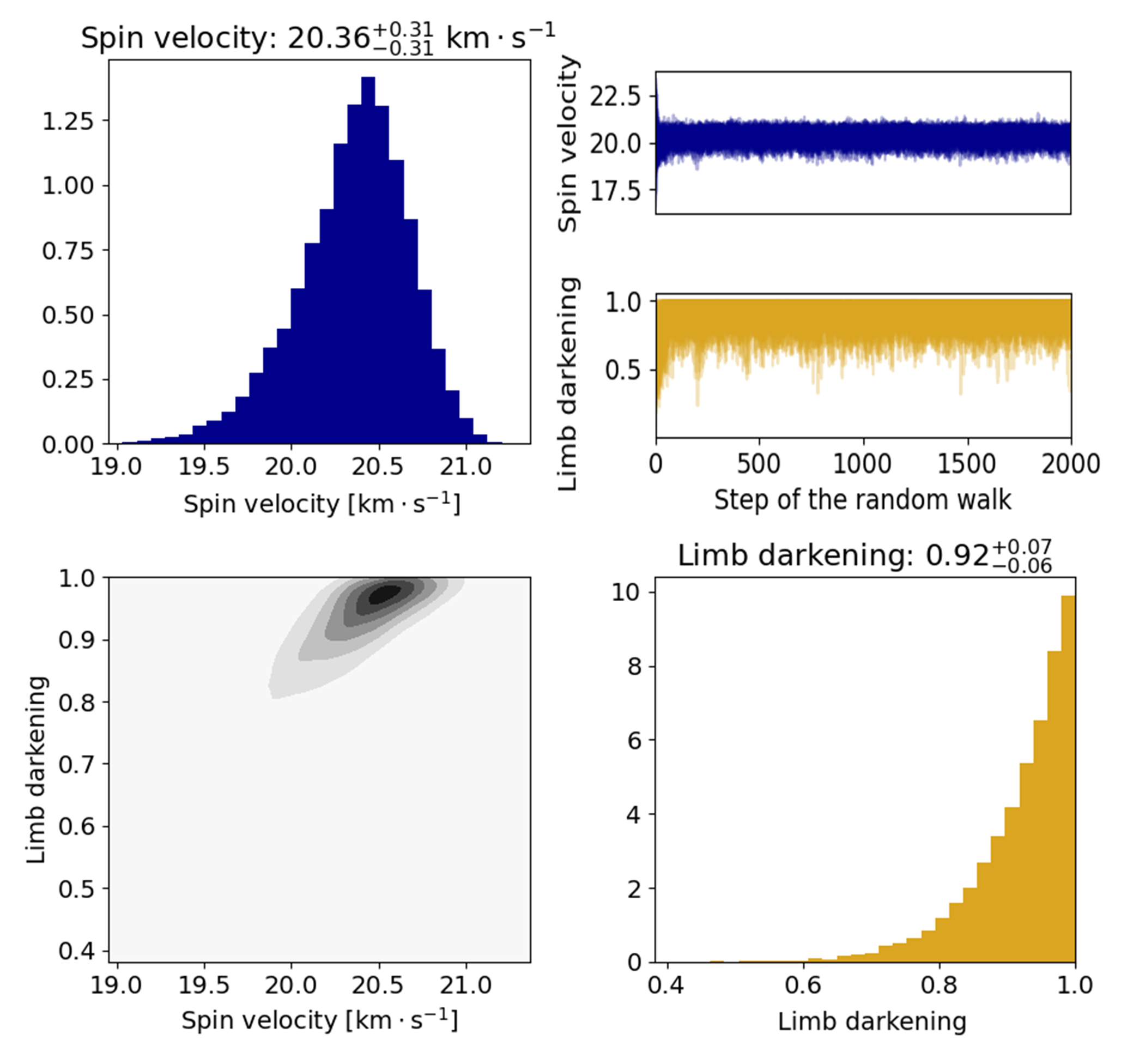}
\caption{Results from MCMC fitting of the spin velocity and limb darkening parameter. Upper left: Histogram for the spin velocity. Upper right: Evolution of the spin velocity (blue) and limb darkening parameter (gold) along the MCMC random walk. Lower left: Correlation between spin velocity and limb darkening. Lower right: Histogram for the limb darkening parameter.}
\label{f:broadening}
\end{figure*}

In order to evaluate the rotational velocity of the planet, we perform a Markov Chain Monte Carlo (MCMC) procedure over a two-dimensional grid of parameters representing a broadening kernel that we apply to the H$_2$O model and compare to the observations. The two parameters are the rotational velocity $v_{\rm rot}$ and the linear limb darkening parameter $c$, where $c = 0$ indicates no limb darkening, and $c = 1$ indicates very strong limb darkening. This is essentially identical to the corresponding procedure in \citet{landman2024}, except that we use a slightly different broadening procedure from \citep{carvalho2023}. This provides a computation-efficient way of accounting for the variation in broadening as function of wavelength, which is particularly important in our case due to the wide (2800--4800 nm) wavelength range that we cover with the observations. The results are shown in Fig.~\ref{f:broadening}. We find a best-fit solution of $v_{\rm rot} = 20.36 \pm 0.31$ km/s and $c = 0.92^{+0.07}_{-0.06}$. These are very high precision constraints for $\beta$~Pic b, consistent with the high significance and low systematic noise of the water feature as discussed above; and fully consistent within error bars with the derived rotational velocities of 19.9$\pm$1.0 km/s from \citet{landman2024} and 22$\pm$2 km/s from \citet{parker2024}. Interestingly, our derived limb darkening parameter, $c = 0.92^{+0.07}_{-0.06}$, is very high and pushing on the upper end of the parameter range, just like in \citet{landman2024} (0.8$\pm$0.2). The limb darkening parameter is included in the fitting primarily to avoid potential systematic errors on the rotational velocity estimation that could arise from omitting it. Hence, we do not draw any firm conclusions on the limb darkening itself here, except to note that the values are arguably surprisingly high, given the long wavelengths of both the K band and L band observations, and the fact that limb darkening is generally expected to decrease with increasing wavelength. For reference, \citet{parker2024} do not fit for a value for the limb darkening parameter, but instead fix it at $c = 0.3$ for their M band observations, based on models of brown dwarf atmospheres. If we strictly enforce $c = 0.3$ in the same way, we acquire a lower rotational velocity of 18.7 km/s. This is still marginally consistent with the \citet{landman2024} and \citet{parker2024} values at the $\sim$2$\sigma$ deviance level, but is a comparatively poor fit to our data, as can be read out from the limb darkening posterior resulting from a uniform prior on $c$ as in Fig. \ref{f:broadening}.

\begin{figure*}[htb]
\centering
\includegraphics[width=18.5cm]{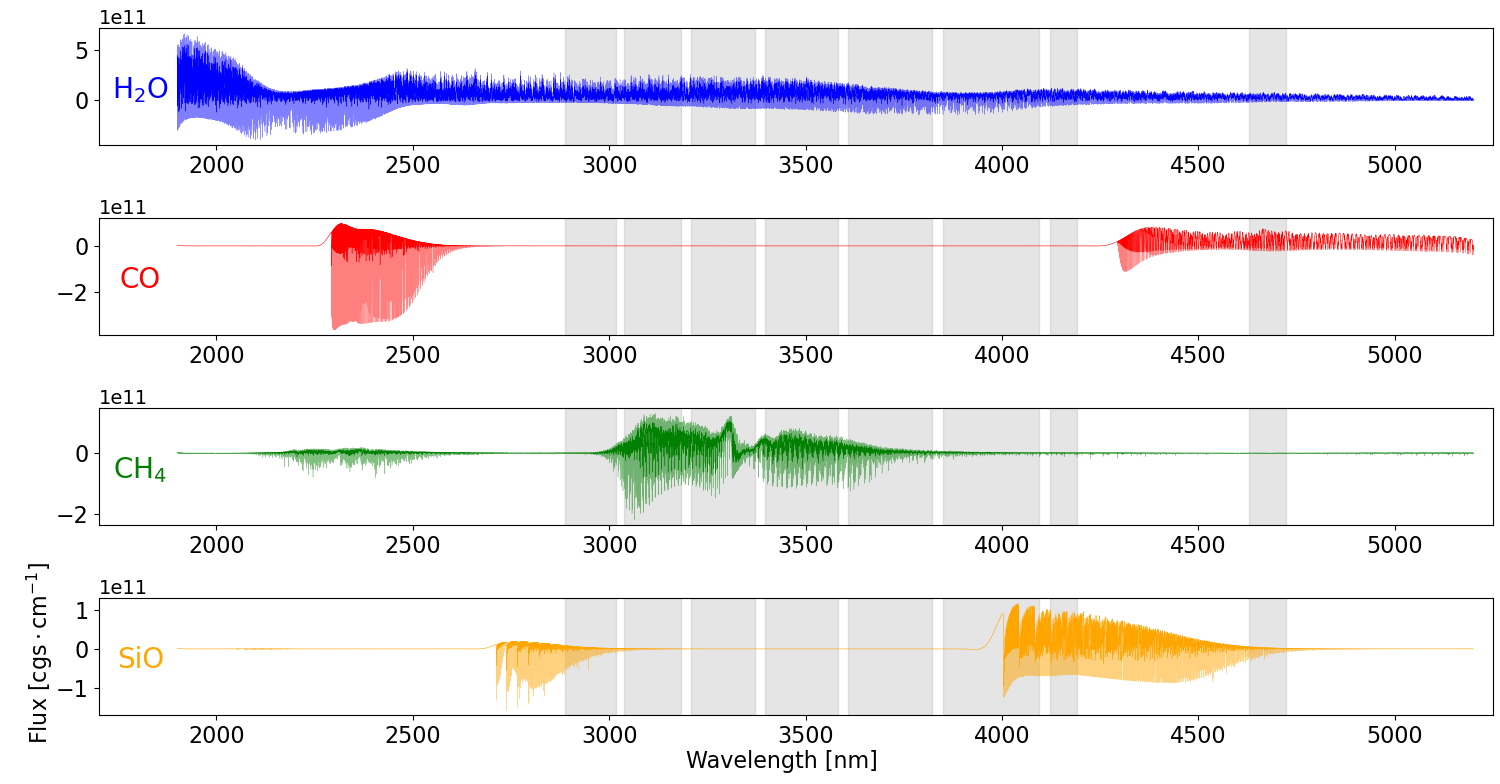}
\caption{Model spectra after continuum subtraction, showing where strong lines occur for four different molecules: H$_2$O in blue, CO in red, CH$_4$ in green, and SiO in orange. The grey regions are the wavelength ranges covered by our observations.}
\label{f:mainmol}
\end{figure*}

In addition to H$_2$O, the only other molecule that has so far been securely detected in $\beta$~Pic b is CO \citep{snellen2014}. Unlike for H$_2$O, which has molecular transitions all across the L band making it a strong region for such detections, CO has essentially no lines at all across the entire L band range (see Fig.~\ref{f:mainmol}). In our data, only a single spectral window in our M band setting covers any CO transitions. We therefore expect the data to be unsuitable for finding CO, and indeed, while our cross-correlation with a CO model does show a peak at the velocity of planet b, it is not at a formally significant level, and we therefore cannot put strong additional constraints on CO beyond what has been done in previous studies in wavelength regions with higher CO opacities \citep[e.g.][]{landman2024,parker2024}. However, the presence of CO lines in the M band window covered by our observation is supported by the fact that cross-correlation with a model that includes both H$_2$O and CO slightly increases the $S/N$ relative to a cross-correlation with a model containing only H$_2$O, going from 15.0 without CO to 15.4 with CO included.

A more promising molecule in terms of opacities in the L band range is SiO. A tentative detection of SiO in the atmosphere of $\beta$~Pic b was reported in \citet{parker2024}, based on cross-correlation with a rotationally broadened SiO model spectrum, giving $S/N = 4.3$, although the tentative nature of the detection was emphasised by the lack of any clear signal in cross-correlations with a non-broadened model spectrum. The spectral setting used in \citet{parker2024} had SiO lines only in a single spectral window. In our observations, SiO lines of similar strength are covered by 2--3 different spectral windows, with a similar integration time depth per window. Hence, if the $S/N = 4.3$ feature in \citet{parker2024} corresponds to a real SiO detection, we should expect to observe such a signature at a $S/N$ in the range of $\sim$6--7 in our data. We therefore performed cross-correlations with SiO models, where we used both models produced by our standard procedure described in Sect.~\ref{s:models}, as well as the exact model used in \citet{parker2024}, kindly provided by L. Parker, with and without rotational broadening. Neither cross-correlation however shows any SiO detection. While there is a bump in the rotationally broadened cross-correlation (see Fig.~\ref{f:injsio}), it is not quite at the right velocity, and it is not statistically significant, with $S/N$ of only 2.2. Given that the previous potential detection was cautioned as being only tentative in \citet{parker2024}, the non-detection at significant levels in this study implies that SiO does not occur at (yet) detectable levels in the photosphere of planet b.   

\begin{figure*}[htb]
\centering
\includegraphics[width=18.5cm]{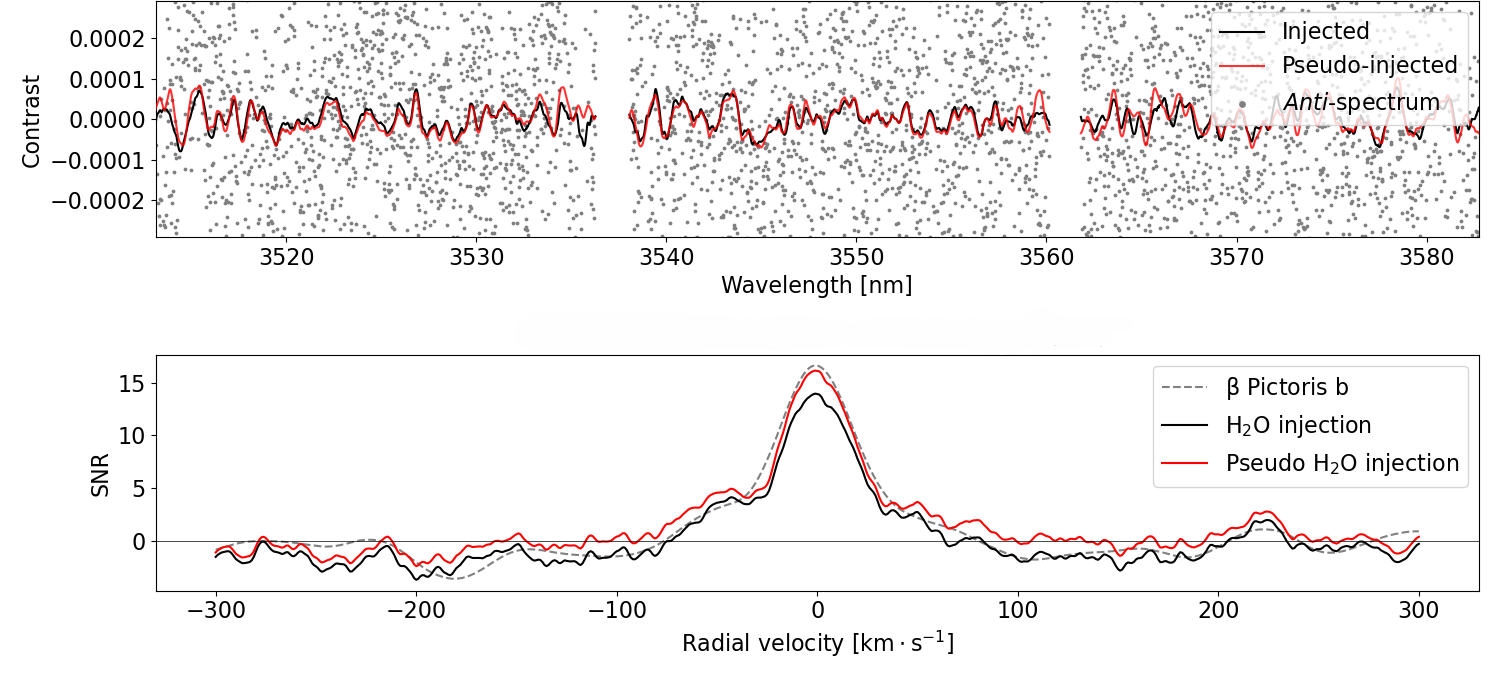}
\caption{Comparison of the signal injection scheme compared to the faster `pseudo-injection' scheme described in Sect.~\ref{s:injection}. Top: Direct comparison of the injected spectra using regular injection (black line) and pseudo-injection (red line). Also shown as grey dots are extracted spectral points at the location of the injection. Bottom: CCF outputs using injection (black line) and pseudo-injection (red line), compared to the actual water CCF for planet b (dashed line). The injection and pseudo-injection are both well representative of the real signal.}
\label{f:pseudo}
\end{figure*}

Many other molecules that have been discussed in the context of planetary atmospheres have good line regions in the L band spectral range, including CH$_4$, CO$_2$, NH$_3$, HCN, C$_2$H$_2$, H$_2$S, O$_3$, and PH$_3$. We have attempted cross-correlations for all of them, but with no significant detections (see Appendix). In order to assess to which extent each molecule would have been detectable if it were present, we have performed (pseudo-)injections for each molecule as described in Sect.~\ref{s:injection}. Fig.~\ref{f:pseudo} demonstrates that the fast pseudo-injection procedure produces results that are equally well representative of expected signal recovery as the slower full injection scheme. By observing at which volume mixing ratios (VMRs) a given molecule is recoverable, we can exclude the presence of each molecule at those VMR levels, while molecular presence at lower levels than the recoverable limit are still possible. In this context, it is important to note that the detectability of a given molecule only increases with increasing VMR up to a point. Once the VMR becomes high enough that a substantial number of individual lines start saturating and blending, the detectability starts decreasing as VMR increases further. Hence, while very high concentrations (say, 1\% VMR) of relatively exotic molecules can probably be seen as physically dubious, they are typically not formally excluded by the spectroscopy. The levels at which each compound is detectable (or not) are summarised in Table~\ref{t:vmr}. Here we use 3$\sigma$ boundaries since, while a 3$\sigma$ level would not be sufficient for a definitive detection, it would be sufficient for a tentative detection worthy of further study, similar to the SiO tentative feature from \citet{parker2024}. 

The injection procedure demonstrates that the $\beta$~Pic~b spectra would have enabled detection of a wide range of molecules, had they been present in suitable amounts. The sensitivity to CO and CO$_2$ is however insufficient for detection, which is as expected, given that both molecules have few lines in the spectral windows covered by the observations. An intuitively more surprising feature is the low sensitivity to CH$_4$, which has lines across the comparatively wide $\sim$3.0--3.6~$\mu$m. The reason for the relatively poor sensitivity to CH$_4$ is a combination of the facts that its lines are still relatively few and shallow compared to other tested molecules; that it is affected by tellurics to a greater extent than many other molecules; and that the CH$_4$ lines reside at relatively short wavelengths within the covered span, where the planetary continuum is fainter, which decreases the achievable $S/N$.

\begin{table}[htb]
\caption{Injection-based detection limits for different molecules in the atmosphere of $\beta$~Pic b.}
\label{t:vmr}
\centering
\begin{tabular}{lll}
\hline
\hline
Molecule & VMR$_{\rm low}$ & VMR$_{\rm high}^{a}$ \\
 & (3$\sigma$) & (3$\sigma$) \\
\hline
H$_2$O  &       $< 1.0 \times 10^{-6}$  &       $1.0 \times 10^{-2}$    \\
CO      &       N/A     &       N/A     \\
CH$_4$  &       N/A     &       N/A     \\
SiO     &       $1.7 \times 10^{-5}$    &       $3.8 \times 10^{-4}$    \\
CO$_2$  &       N/A     &       N/A     \\
NH$_3$  &       $1.9 \times 10^{-5}$    &       $> 1.0 \times 10^{-1}$  \\
HCN     &       $1.7 \times 10^{-6}$    &       $1.9 \times 10^{-4}$    \\
C$_2$H$_2$      &       $8.7 \times 10^{-5}$    &       $3.8 \times 10^{-2}$    \\
H$_2$S  &       $6.3 \times 10^{-5}$    &       $> 1.0 \times 10^{-1}$  \\
O$_3$   &       $8.3 \times 10^{-5}$    &       $> 1.0 \times 10^{-1}$  \\
PH$_3$  &       $7.7 \times 10^{-4}$    &       $5.5 \times 10^{-3}$    \\
\hline
\end{tabular}
\begin{list}{}{}
\item[$^{\mathrm{a}}$] VMR$_{\rm low}$: Lowest marginally detectable volume mixing ratio. VMR$_{\rm high}$: Highest marginally detectable volume mixing ratio. N/A indicates that the molecule is undetectable at any VMR.
\end{list}
\end{table}

\subsection{Planet c}
\label{s:planetc}

A particularly beneficial aspect of the L band for planet spectroscopy purposes is the relatively low intrinsic contrast between star and planet, for many types of planets. In K band, $\beta$~Pic c would be very much harder to detect with CRIRES+ than $\beta$~Pic b, primarily because the noise from the stellar PSF is much higher for c than for b due to the smaller angular separation. This is true also in the L band, but the difference is much smaller, since the balance between background noise versus PSF noise leans much more in the direction of background dominance in the L band. The difference in noise level between the positions of planet c (at maximum separation) and planet b in the CRIRES+ data is only a factor $\sim$3 at the long-wavelength end and a factor $\sim$7 at the short-wavelength end. Hence, we considered it worthwhile to attempt detection of molecular features in the atmosphere of planet c, since it was anyway in the slit that was aligned for including planet b, thanks to the edge-on orientation of the system. 

Unfortunately, the timing of the observations was far from ideal for studying planet c. In Nov-Dec 2022, its projected separation from the star was $<$50 mas (and therefore $<$1 pixel), which is not a workable configuration. In Jan-Feb 2023, the separation was 50--80 mas, which is also not ideal. In Sep-Nov, on the other hand, the separation was essentially perfect at close to 150 mas -- however, at this particular orbital configuration (south-western quadrature), it happens to be the case that the systemic RV of the $\beta$~Pic system and the orbital RV of planet c around its star cancel out almost exactly. The shift between the planet c atmospheric lines and the corresponding telluric lines is then determined almost entirely by the projected barycentric velocity of Earth around the Sun, which is small in the case of $\beta$~Pic (6--7 km/s in late Autumn 2023) since it is located far from the ecliptic plane. For future efforts, the ideal ephemeris for ground-based high-resolution spectroscopy of $\beta$~Pic c is when it resides near its north-eastern quadrature, where it is simultaneously at a large angular separation from its star, and at a high velocity shift relative to telluric lines. 

\begin{figure*}[htb]
\centering
\includegraphics[width=18.5cm]{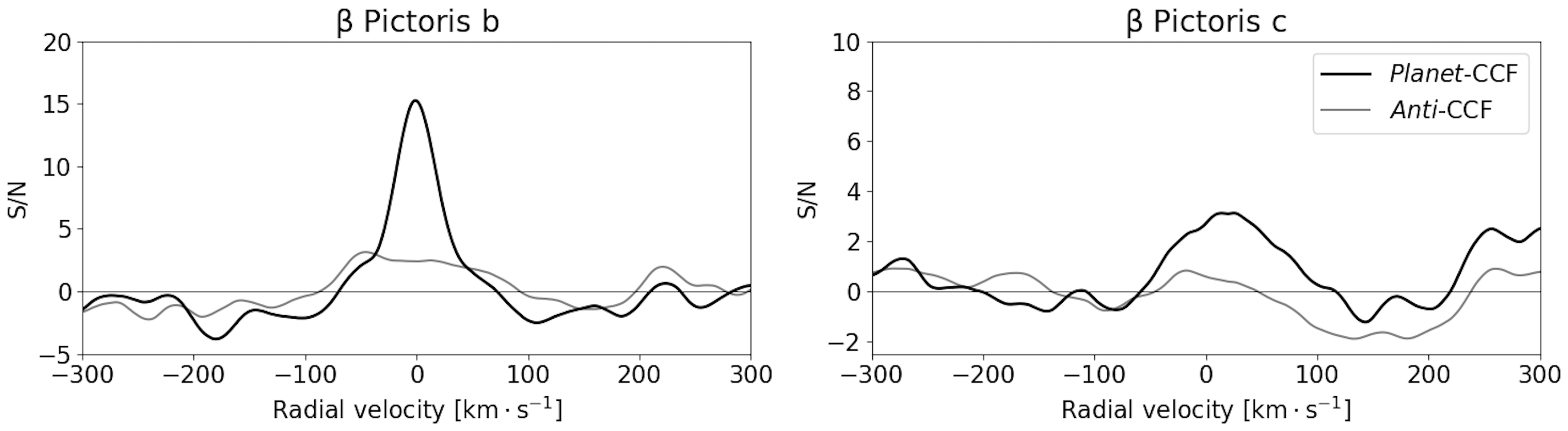}
\caption{Cross-correlation functions of $\beta$~Pic b (left) and c (right). The black lines are the cross-correlation functions at the locations of the planets, while the grey lines are the corresponding functions at the exact opposite side of the star relative to the planets. Note the different scaling on the y-axes between the two panels. Planet b is clearly detected, while planet c shows a bump that can only be seen as a tentative feature.}
\label{f:waterccf}
\end{figure*}

For the cross-correlation of a water model with the extracted spectrum of planet c, we use only the epochs from Jan 2023 onwards when the planet resides at a reasonable separation from its parent star. The result is shown in Fig.~\ref{f:waterccf}. Interestingly, a peak is seen at a significance level of $S/N = 3.14$. The peak is an intriguing feature, although several caveats should be kept in mind for its interpretation. Firstly, the significance level is well below $S/N = 5$, which we regard as a threshold required for a definitive detection. Secondly, it is not clear that planet c could yield a significantly strong H$_2$O feature to reach $S/N = 3.14$ in the data. Injection tests cannot be done in the same way as for planet b, or at least not to the same level of rigor, because for planet c we have much fewer constraints on its absolute brightness level. If planet c had a similar brightness level as planet b, $S/N = 3.14$ might be quite reasonable. This would be consistent with the fact that the planets have similar masses \citep{lacour2021}, and presumably, near-equal ages. However, GRAVITY spectra \citep{nowak2020} imply that planet c is substantially colder than planet b, as well as smaller. In that context, the predicted line strengths are far weaker, and no significant or near-significant detection would be expected. 

For these reasons, we emphasise that the peak in the water cross-correlation for planet c should only be regarded as a tentative feature, requiring further data to confirm its status. We have also tried cross-correlations with the other molecules listed in Sect. \ref{s:planetb}, but without any detections - which is entirely as expected, given the lower signal-to-noise ratios relative to the case of planet b.

\section{Discussion}
\label{s:discussion}

\begin{figure*}[htb]
\centering
\includegraphics[width=18.5cm]{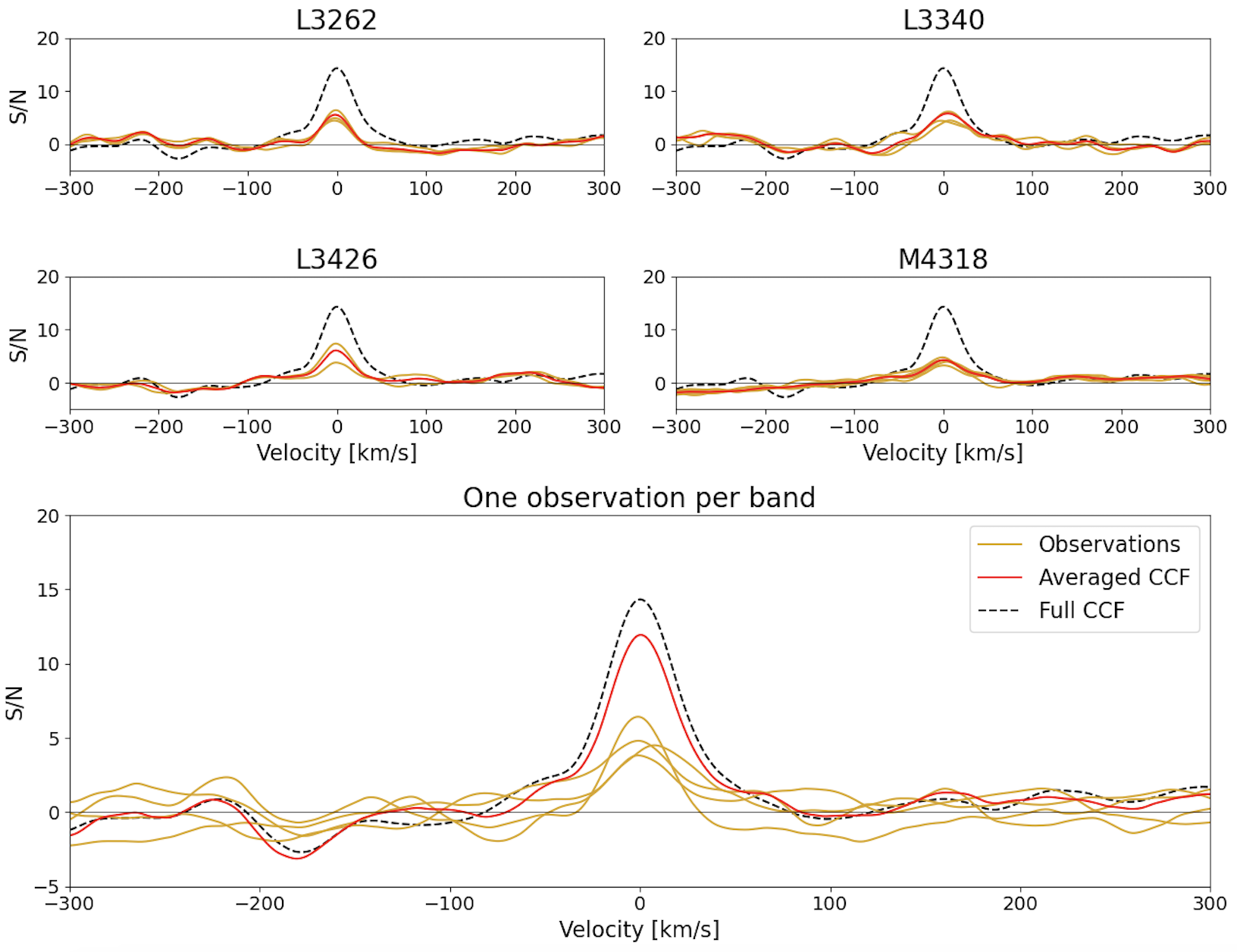}
\caption{Cross-correlation results (normalised to units of $S/N$) for different ways of combining the individual data sets. In each case, the black dashed line represents the result when combining all the data together. Top left: Orange lines are the CCFs from each individual observation taken in the L3262 setting. The red line is the combination of all L3262 observations. No dramatic improvement in $S/N$ is acquired from this combination. Top right: Same, but for the case of L3340. Middle left: Same, for the case of L3426. Middle right: Same, for the case of M4318. Bottom: Here, a set of single observations from each spectral setting is plotted as orange lines. Their combination is plotted as a red line. In this case, a significant improvement is gained from combining the data sets, almost reaching the $S/N$ level from the full data set.}
\label{f:combining}
\end{figure*}

Thanks to the depth and spectral coverage of the L and M band data in this study, and the strength of the water feature in planet b, it is possible to draw conclusions regarding which observational parameters are the most important for maximising sensitivity in future studies. We illustrate this in Fig.~\ref{f:combining}. The H$_2$O line is detected at a significance level of around $S/N = 5$ in every individual setting at every epoch. When combining different epochs for the same setting, the $S/N$ barely increases at all. By stark contrast, when different spectral settings are combined, the $S/N$ increases rapidly. This implies that, despite the observations being largely thermal background limited for planet b, noise sources that are systematic in the time domain (but variable in the wavelength domain) dominate the cross-correlation $S/N$. The effect is closely connected to the spectrally non-white oscillations along the baseline of the cross-correlations, which decrease relative to the peak when different spectral settings are combined. We recall that previous spectroscopic analyses based on single spectroscopic settings often show large oscillations of this nature \citep[e.g.][]{landman2024}, supporting this interpretation. 

Residual features of telluric absorption could cause oscillations in the cross-correlation functions, although since the different epochs are taken over a wide time span, the telluric features move around in velocity space relative to the planet. They should therefore average down when combining different epochs of the same setting, which is not observed. Probable contributing effects to the oscillations are inaccuracies in the water models (for example, the line lists) or regularities in the spacings between individual water lines, causing partial degeneracies in the cross-correlation output. Whatever the underlying cause, these results have implications for cross-correlation studies of planetary atmospheres in general, implying that an optimal observational approach for maximising $S/N$ is to spread the wavelength coverage as much as practically possible. In other words, if several spectral windows are available in which a planet is expected to be observable and have interesting molecular lines, then multiple separate shallow observations covering all of those spectral windows is likely to be a more fruitful strategy than spending the corresponding observing time on deep observations of a single window. These results also imply that in a future study, there could be potentially much to gain from combining all of the existing high-resolution data sets of $\beta$~Pic b, covering K band, L band and M band, for maximising $S/N$ and potentially detect a wider range of molecules in the atmosphere. 

The main advantages of the L band range for high-resolution spectroscopy, aside from sensitivity to a wide range of molecules, are the favourable planet-to-star contrast for a wide range of planetary temperatures, and the relative ease to acquire high adaptive optics performance and PSF stability. The main disadvantage, meanwhile, is the high level of thermal background. This limits the utility of 8m-class telescopes and their instruments (such as VLT/CRIRES+) for L band exoplanet spectroscopy. VLT/CRIRES+ is excellently suited for studying $\beta$~Pic since the $\beta$~Pic planets are the brightest planets known, thanks to their proximity, youth, and high masses. Fainter planets will increasingly become dominated by the thermal noise, generally necessitating impractically long integration times with CRIRES+. Instead, the major application of L band high-resolution spectroscopy will come with the next generation of large-aperture (20--40 m diameter) telescopes. Such telescopes feature an enormously improved background-limited performance, due to the strong dependence of telescope sensitivity on aperture size. The largest, and therefore least background limited, of such telescopes will be ESO's 39 m Extremely Large Telescope (ELT). One of the first-generation instruments on ELT will be METIS \citep{brandl2021}, a mid-infrared imager and spectrograph with a spectral resolution in the L and M bands of up to $\sim$100\,000. Thanks to the beneficial properties of the L band range once the background limit is improved, METIS will be an ideal instrument for detailed characterisation of a large number of directly imaged planets. Similar instruments with L band high-resolution capability are foreseen also for the other next-generation facilities: The Giant Magellan Telescope Near-Infrared Spectrograph \citep[GMTNIRS; ][]{jaffe2016} on the 20 m Giant Magellan Telescope, and the Mid-IR Camera, High-disperser \& IFU spectrograph \citep[MICHI; ][]{packham2018} on the 30 m Thirty Meter Telescope (TMT).

\section{Conclusions}
\label{s:summary}

In this study, we have analysed the results of CRIRES+ LM band spectroscopy of the $\beta$~Pic system, primarily for the purpose of characterising $\beta$~Pic b, and secondarily for searching for detectable features in the atmosphere of $\beta$~Pic c. We outline our main conclusions below:

\begin{enumerate}
\item We detected H$_2$O in the atmosphere of $\beta$~Pic b at a significance level of $S/N$ = 15.0. If CO is included in the model used for the cross-correlation, the $S/N$ increases to 15.4. The rotational velocity was measured with a high precision as $v_{\rm rot} = 20.36 \pm 0.31$. These results are well consistent with previous studies \citep{landman2024,parker2024}.
\item We do not confirm a previous tentative indication of SiO \citep{parker2024} in the atmosphere of $\beta$~Pic b. A mild peak appears in the CCF, but it is not at the expected velocity, and is not statistically significant at $S/N = 2.2$. Based on the theoretical SiO spectrum, our sensitivity should be the highest acquired so far. Still, injection results imply that detection of actual SiO should be challenging even in our data set. 
\item We detected a peak in the CCF for H$_2$O at the expected separation and velocity of $\beta$~Pic c. However, it is not a statistically significant detection ($S/N = 3.14$). It is also unclear if planet c is intrinsically bright enough to cause the feature. We therefore consider the feature as tentative. Further studies would be necessary in order to establish whether or not the feature is real.
\item Combining data sets with different spectral settings rapidly increases $S/N$ for the H$_2$O feature of planet b, while combining data sets with equal spectral settings results only in small improvements at best. This implies that the widest possible spectral coverage will be a more important factor than the deepest possible integrations when planning CCF-based exoplanet studies. 
\end{enumerate}

Altogether, the observations demonstrate the power of high-resolution spectroscopy in the LM band range. This wavelength range is currently limited for exoplanet purposes by high thermal background, but will greatly expand in applicability with the advent of next-generation telescopes and instrumentation, including ELT/METIS \citep{brandl2021}, GMT/GMTNIRS \citep{jaffe2016}, and TMT/MICHI \citep{packham2018}.

\begin{acknowledgements}
M.J. gratefully acknowledges funding from the Knut and Alice Wallenberg Foundation and the Swedish Research Council. The authors thank Thomas Marquart, Luke Parker, Rico Landman and Jayne Birkby for useful discussions regarding the analysis of CRIRES+ data. This work has made use of data from the European Space Agency (ESA) mission {\it Gaia} (\url{https://www.cosmos.esa.int/gaia}), processed by the {\it Gaia} Data Processing and Analysis Consortium (DPAC, \url{https://www.cosmos.esa.int/web/gaia/dpac/consortium}). Funding for the DPAC has been provided by national institutions, in particular the institutions participating in the {\it Gaia} Multilateral Agreement. The study has made use of the CDS, NASA ADS, and NASA exoplanet archive services, and python packages matplotlib \citep{hunter2007} and numpy \citep{harris2020}.
\end{acknowledgements}

\begin{appendix}

\onecolumn

\section{Additional molecular lines}

\begin{figure*}[htb]
\centering
\includegraphics[width=18.5cm]{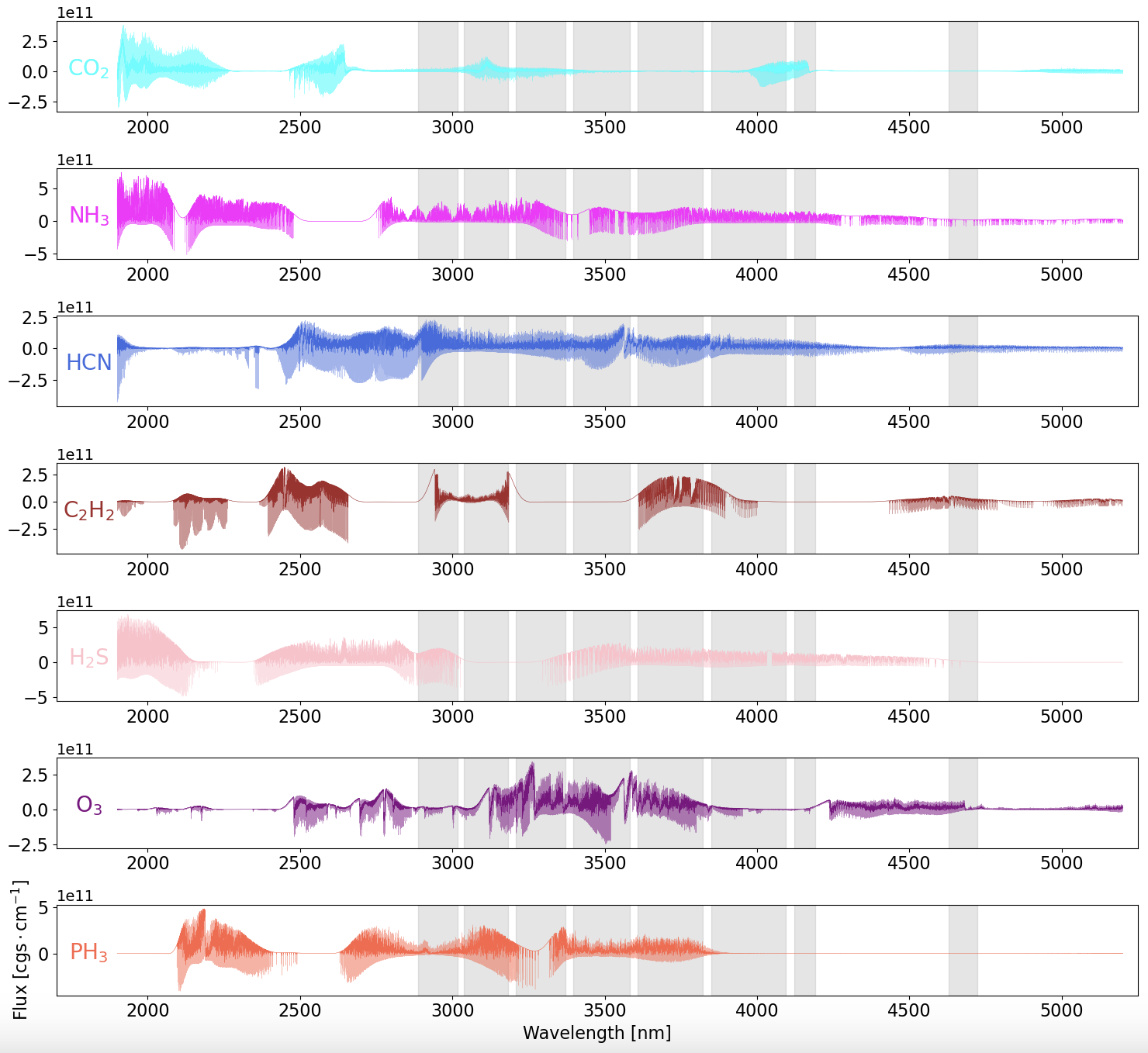}
\caption{Theoretical spectra after continuum subtraction of molecules that we have searched for in the cross-correlation analysis. A similar figure for H$_2$O, CO, CH$_4$ and SiO is shown in the main text, Fig.~\ref{f:mainmol}. From top to bottom, the additional molecules shown here are CO$_2$ in cyan, NH$_3$ in magenta, HCN in light blue, C$_2$H$_2$ in dark red, H$_2$S in pink, O$_3$ in purple, and PH$_3$ in orange. Grey areas correspond to the wavelength windows covered by our observations.}
\label{f:extramol}
\end{figure*}

Fig.~\ref{f:extramol} shows the expected lines in the L band region (and its surroundings) for 7 molecules that have been included in the CCF analysis but not discussed at depth in the main text.


\section{H$_2$O injection}

\begin{figure*}[htb]
\centering
\includegraphics[width=18.5cm]{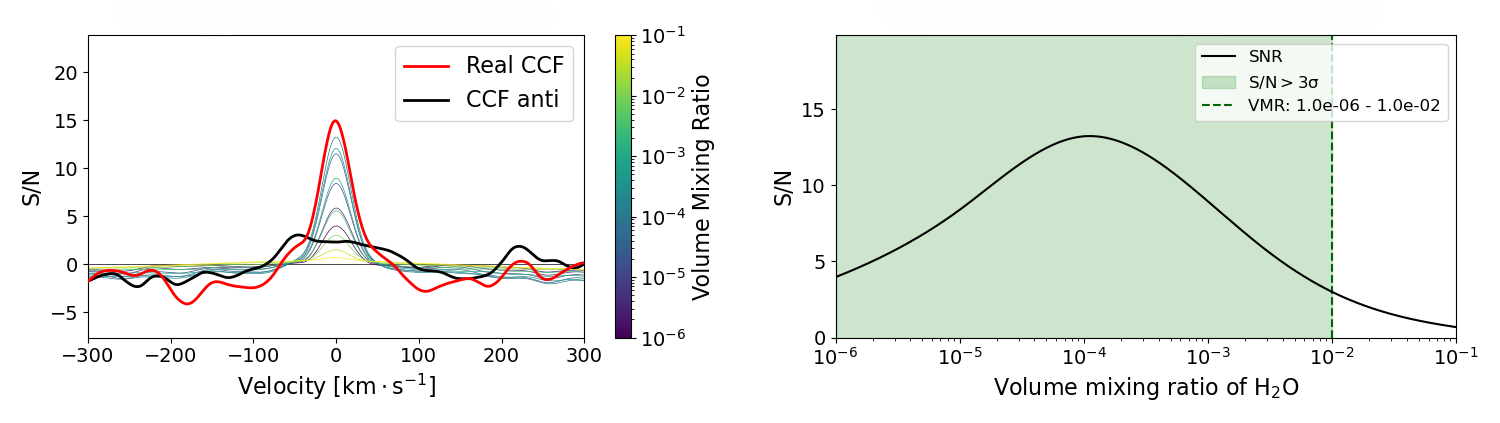}
\caption{Left: CCFs in units of $S/N$ for H$_2$O in the atmosphere of $\beta$~Pic b. Thick red line: Actual CCF for the location of the planet. Thick black line: CCF for the location at the opposite side of the star. Thin lines: CCF for injected H$_2$O, colour coded by different volume mixing ratios as shown in the colour bar. Right: $S/N$ of injected signals as function of VMR. The green shaded area shows at which range of VMRs the molecule would have been marginally detectable.}
\label{f:injh2o}
\end{figure*}

Fig.~\ref{f:injh2o} shows the results of the injection analysis for H$_2$O. The injection procedure predicts strong detectability of H$_2$O for a wide range of VMRs, consistent with the actual detection in the CCF at the location of the planet.

\FloatBarrier

\section{CO injection}

\begin{figure*}[htb]
\centering
\includegraphics[width=18.5cm]{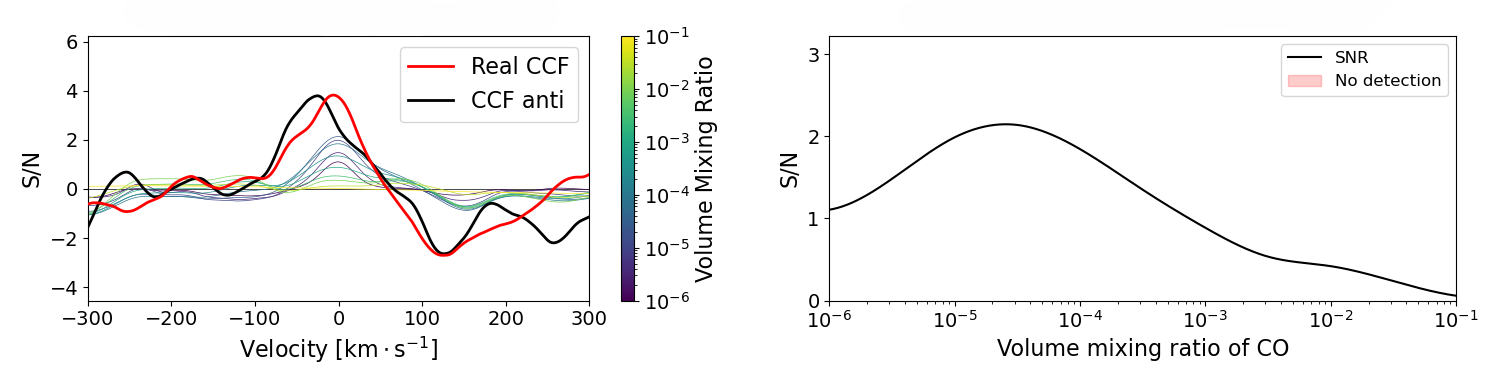}
\caption{Same as Fig.~\ref{f:injh2o}, for CO.}
\label{f:injco}
\end{figure*}

Fig.~\ref{f:injco} shows the results of the injection analysis for CO.

\FloatBarrier

\section{SiO injection}

\begin{figure*}[htb]
\centering
\includegraphics[width=18.5cm]{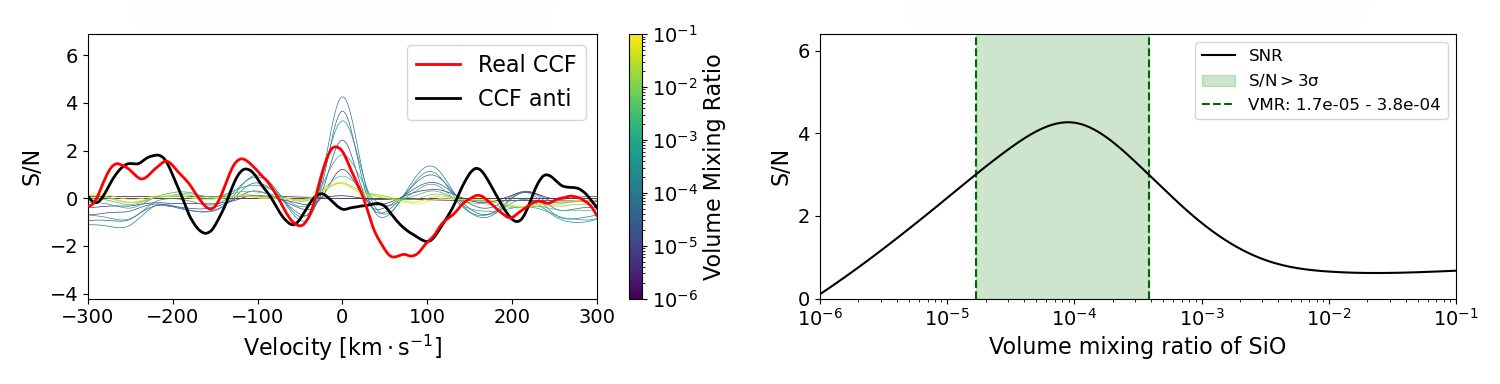}
\caption{Same as Fig.~\ref{f:injh2o}, for SiO.}
\label{f:injsio}
\end{figure*}

Fig.~\ref{f:injsio} shows the results of the injection analysis for SiO.

\FloatBarrier

\section{CH$_4$ injection}

\begin{figure*}[htb]
\centering
\includegraphics[width=18.5cm]{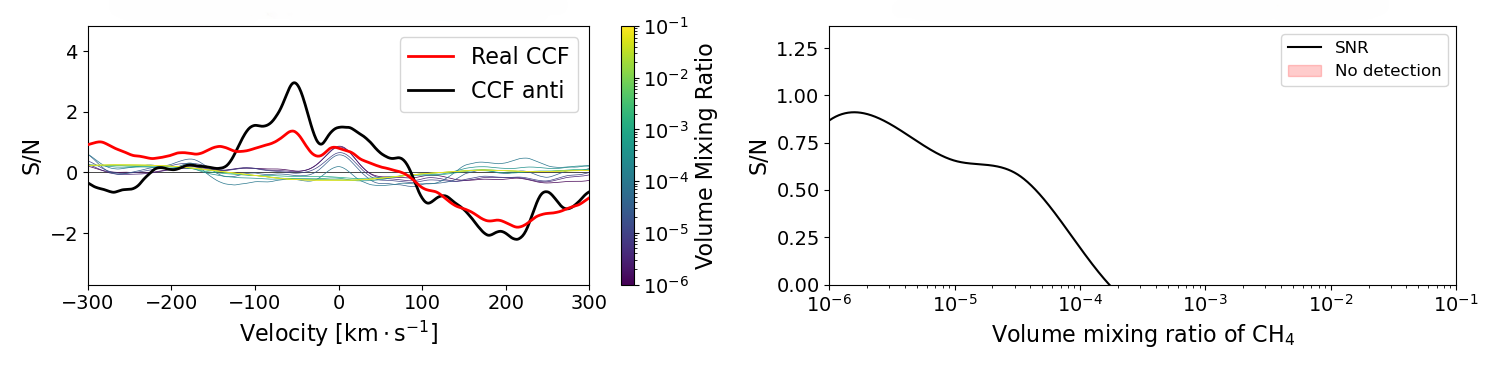}
\caption{Same as Fig.~\ref{f:injh2o}, for CH$_4$.}
\label{f:injch4}
\end{figure*}

Fig.~\ref{f:injch4} shows the results of the injection analysis for CH$_4$.

\FloatBarrier

\section{CO$_2$ injection}

\begin{figure*}[htb]
\centering
\includegraphics[width=18.5cm]{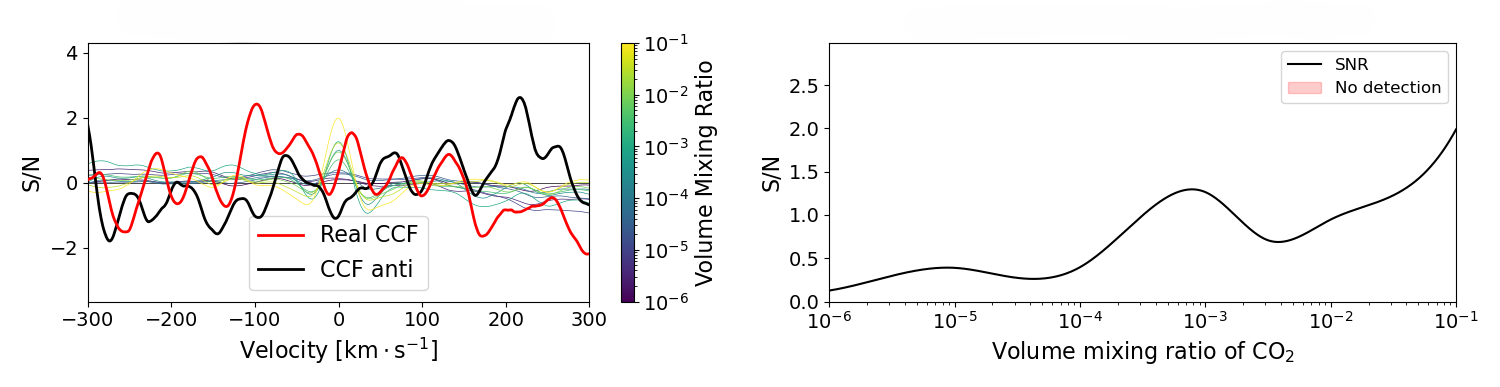}
\caption{Same as Fig.~\ref{f:injh2o}, for CO$_2$.}
\label{f:injco2}
\end{figure*}

Fig.~\ref{f:injco2} shows the results of the injection analysis for CO$_2$.

\FloatBarrier

\section{NH$_3$ injection}

\begin{figure*}[htb]
\centering
\includegraphics[width=18.5cm]{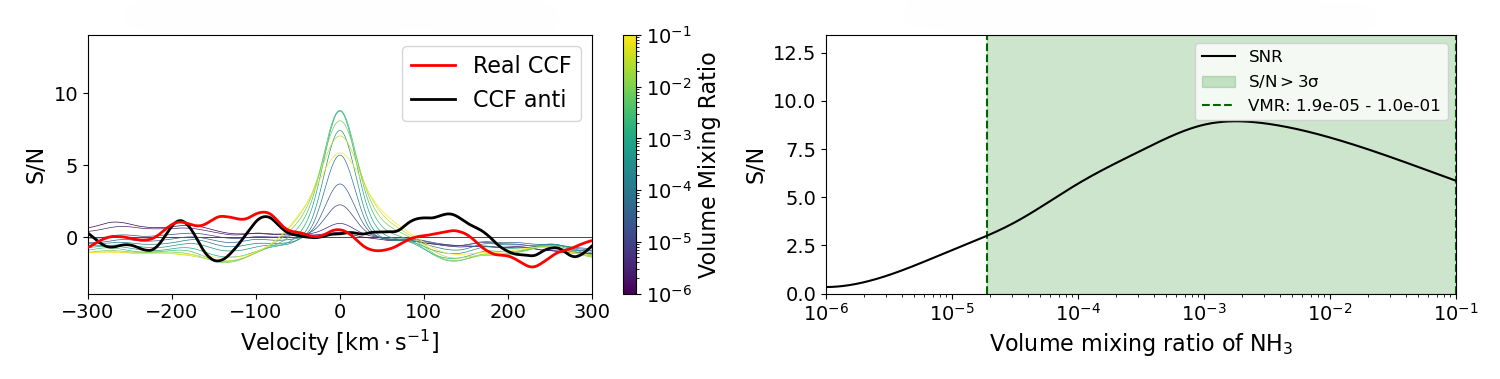}
\caption{Same as Fig.~\ref{f:injh2o}, for NH$_3$.}
\label{f:injnh3}
\end{figure*}

Fig.~\ref{f:injnh3} shows the results of the injection analysis for NH$_3$.

\FloatBarrier

\section{HCN injection}

\begin{figure*}[htb]
\centering
\includegraphics[width=18.5cm]{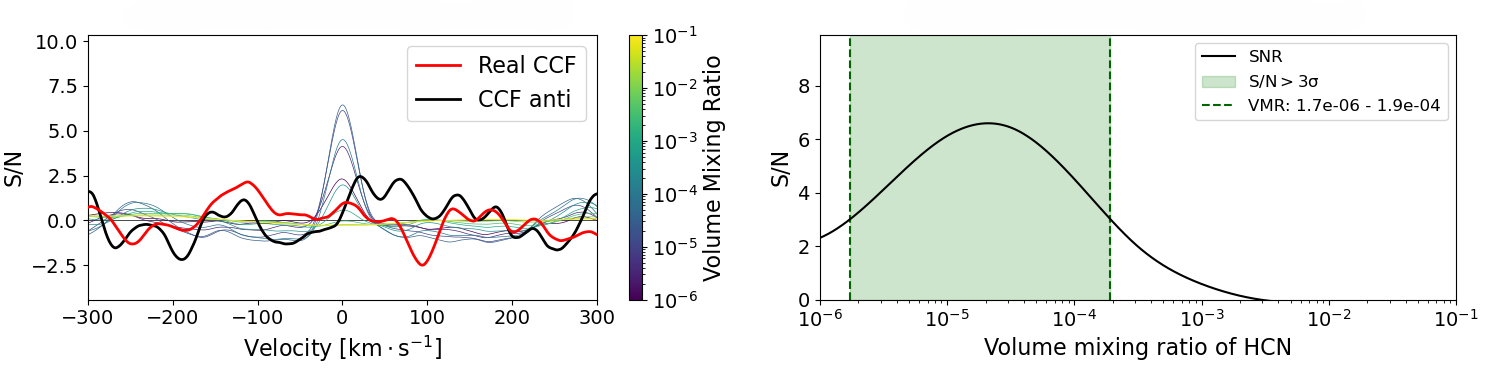}
\caption{Same as Fig.~\ref{f:injh2o}, for HCN.}
\label{f:injhcn}
\end{figure*}

Fig.~\ref{f:injhcn} shows the results of the injection analysis for HCN.

\FloatBarrier

\section{C$_2$H$_2$ injection}

\begin{figure*}[htb]
\centering
\includegraphics[width=18.5cm]{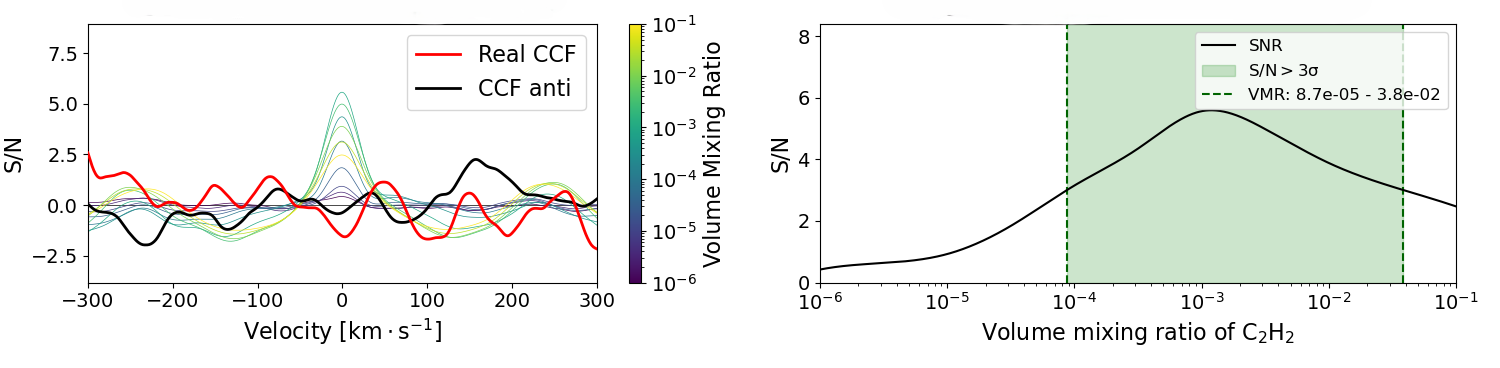}
\caption{Same as Fig.~\ref{f:injh2o}, for C$_2$H$_2$.}
\label{f:injc2h2}
\end{figure*}

Fig.~\ref{f:injc2h2} shows the results of the injection analysis for C$_2$H$_2$.

\FloatBarrier

\section{H$_2$S injection}

\begin{figure*}[htb]
\centering
\includegraphics[width=18.5cm]{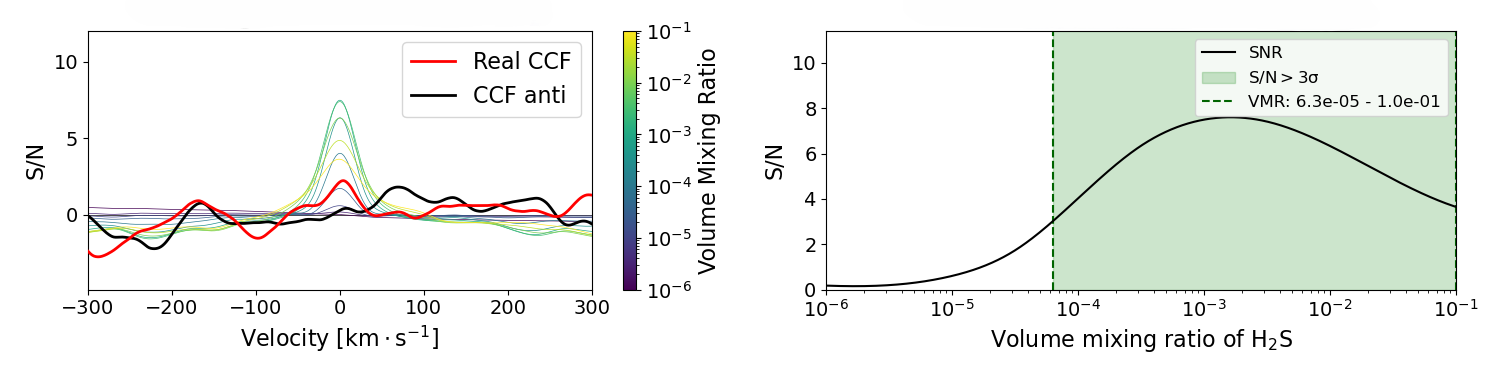}
\caption{Same as Fig.~\ref{f:injh2o}, for H$_2$S.}
\label{f:injh2s}
\end{figure*}

Fig.~\ref{f:injh2s} shows the results of the injection analysis for H$_2$S.

\FloatBarrier

\section{O$_3$ injection}

\begin{figure*}[htb]
\centering
\includegraphics[width=18.5cm]{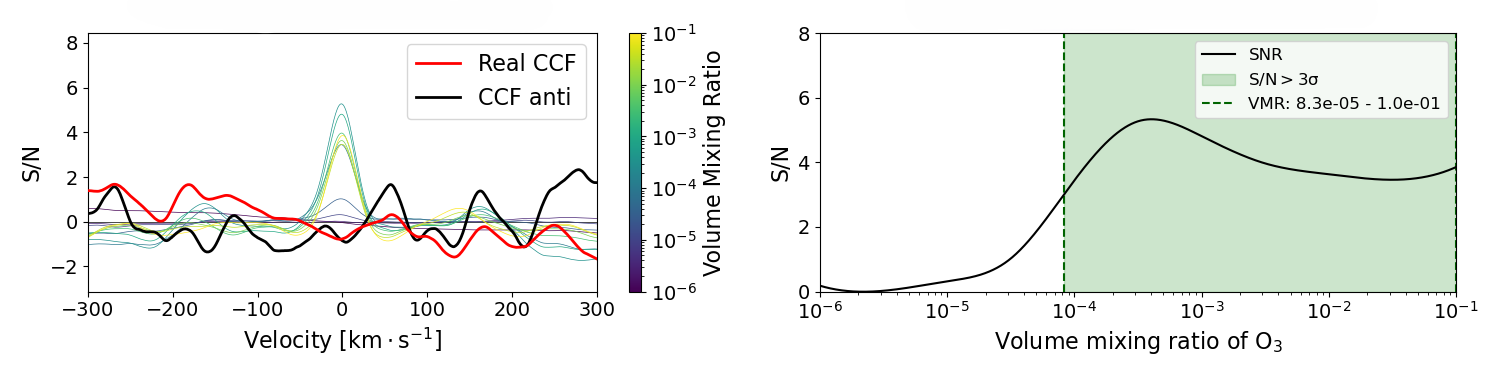}
\caption{Same as Fig.~\ref{f:injh2o}, for O$_3$.}
\label{f:injo3}
\end{figure*}

Fig.~\ref{f:injo3} shows the results of the injection analysis for O$_3$.

\FloatBarrier

\section{PH$_3$ injection}

\begin{figure*}[htb]
\centering
\includegraphics[width=18.5cm]{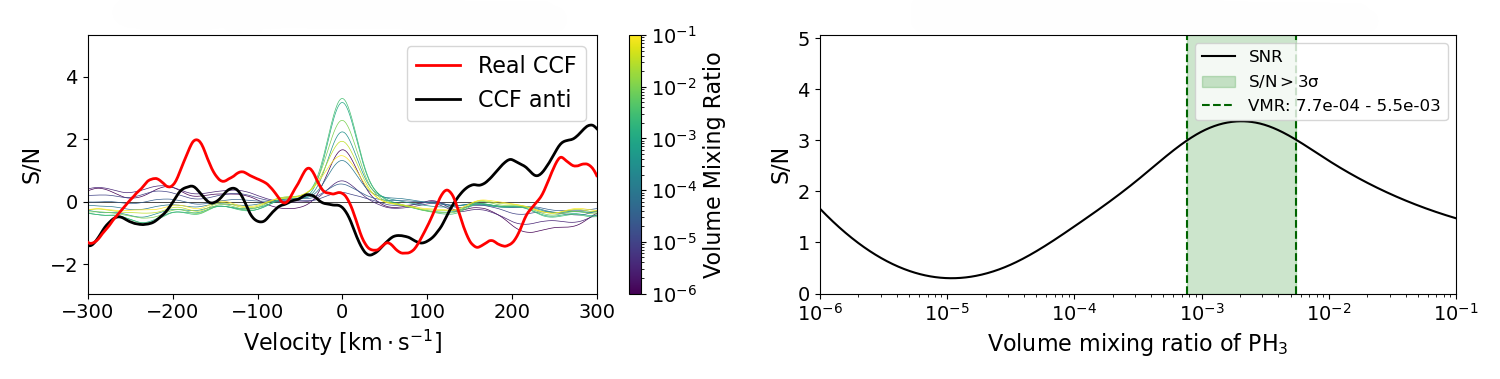}
\caption{Same as Fig.~\ref{f:injh2o}, for PH$_3$.}
\label{f:injph3}
\end{figure*}

Fig.~\ref{f:injph3} shows the results of the injection analysis for PH$_3$.

\end{appendix}

\end{document}